\documentclass[%
preprint,
showpacs,
amsmath,amssymb,
aps,prl,
]{revtex4-1}

\usepackage{graphicx}
\usepackage{dcolumn}
\usepackage{bm}
\usepackage{subfigure}
\usepackage{color}

\begin{document}
\hyphenation{eq-ua-tions diff-er-ent only sce-nario also however equi-librium fila-ment results works re-mains two still account University} 
\title {Scaling regimes in rapidly rotating thermal convection at extreme Rayleigh numbers}
\author{Jiaxing Song$^{1}$}
\author{Olga Shishkina$^{2}$}
\author{Xiaojue Zhu$^{1}$}
\email{zhux@mps.mpg.de}

\affiliation{$^1$Max Planck Institute for Solar System Research, G\"{o}ttingen, 37077, Germany\\
$^2$Max Planck Institute for Dynamics and Self-Organization, G\"{o}ttingen, 37077, Germany}

\begin{abstract}
The geostrophic turbulence in rapidly rotating thermal convection exhibits characteristics shared by many highly turbulent geophysical and astrophysical flows. In this regime, the convective length and velocity scales, heat flux, and kinetic and thermal dissipation rates are all diffusion-free, meaning that they are independent of the viscosity and thermal diffusivity. 
Our direct numerical simulations (DNS) of rotating Rayleigh--B\'enard convection in domains with no-slip top and bottom and periodic lateral boundary conditions for a fluid with the Prandtl number $Pr=1$ and extreme buoyancy and rotation parameters (the Rayleigh number up to $Ra=3\times10^{13}$ and the Ekman number down to $Ek=5\times10^{-9}$) indeed demonstrate these diffusion-free scaling relations, in particular, that the dimensionless convective heat transport scales with the supercriticality parameter $\widetilde{Ra}\equiv Ra\,Ek^{4/3}$ as $Nu-1\propto \widetilde{Ra}^{3/2}$, where $Nu$ is the Nusselt number. 
We further derive and verify in the DNS that with the decreasing $\widetilde{Ra}$ the geostrophic turbulence regime undergoes a transition into another geostrophic regime where the convective heat transport scales as $Nu-1\propto \widetilde{Ra}^{3}$. 
\end{abstract}

\maketitle

Turbulent rotating convection \cite{Ecke23, Kunnen21} is a fundamental mechanism that drives the heat and momentum transport in planets \cite{Busse76, Heimpel05, Ahlers09, Wicht19, Yadav20a, Yadav20b}, as well as the energy source for planetary and stellar magnetic fields \citep{Jones11, Aurnou15, Cheng15, Hanasoge16, Yadav16, Cheng18, Guervilly19, Schumacher20, Tobias21, Landeau22}. 
The parameters of the astrophysical and geophysical flows are too extreme to be realized nowadays in lab experiments and direct numerical simulations (DNS). 
For example, in the Earth's core, the Ekman number $Ek\equiv\nu/(2\Omega L^2)$, which is inverse of the dimensionless rotating rate, can be as low as $10^{-15}$, and the Reynolds number $Re\equiv uL/\nu$, which is the dimensionless flow velocity, can be as high as $10^9$ \citep{Aurnou15, Plumley19, Landeau22}. 
Here $\nu$ is the kinematic viscosity, $\Omega$ the rotating angular velocity, $u$ the characteristic velocity, and $L$ the domain length scale.
To estimate the heat and momentum transport in a particular geophysical or astrophysical system, one needs, first, the scaling relations that hold in the corresponding flow regime and, second, measurements or simulations for a certain range of control parameters, which are not as extreme as in the considered geophysical or astrophysical system, but which anyway belong to the same scaling regime as the considered system. As soon as both objectives are achieved, the results from the labs and supercomputers can be upscaled to the geophysical and astrophysical conditions.

Rotating Rayleigh--B\'enard convection (RRBC) \citep{Ecke23, Zhong09, King09, Stevens09, Stevens10, Stevens13, Ecke14, Stellmach14, Horn2015, Weiss16, Plumley17, Zhang20, Aurnou20, Cheng20, Guzman20, Jiang20, Kunnen21, Bouillaut21, Lu21} is the most studied setup of rotating thermal convection. 
Here, a container of height $L$ and a temperature difference $\Delta$ between its bottom and top is rotated with an angular velocity $\Omega$ about its centrally located vertical axis. 
The main control parameters of the system are $Ek$, the Rayleigh number $Ra\equiv\alpha_T g L^3 \Delta/(\kappa \nu)$, which is the dimensionless temperature difference across the domain, and the Prandtl number $Pr\equiv\nu/\kappa$, a material property.
Here, $\alpha_T$ is the thermal expansion coefficient, $g$ the gravitational acceleration, $\kappa$ the thermal diffusivity.   
The main dimensionless response characteristics 
are $Re$ and the Nusselt number $Nu$ which is the total vertical heat flux normalized by the purely conductive counterpart.

\begin{figure*}[t]\centering\includegraphics[width=1\linewidth]{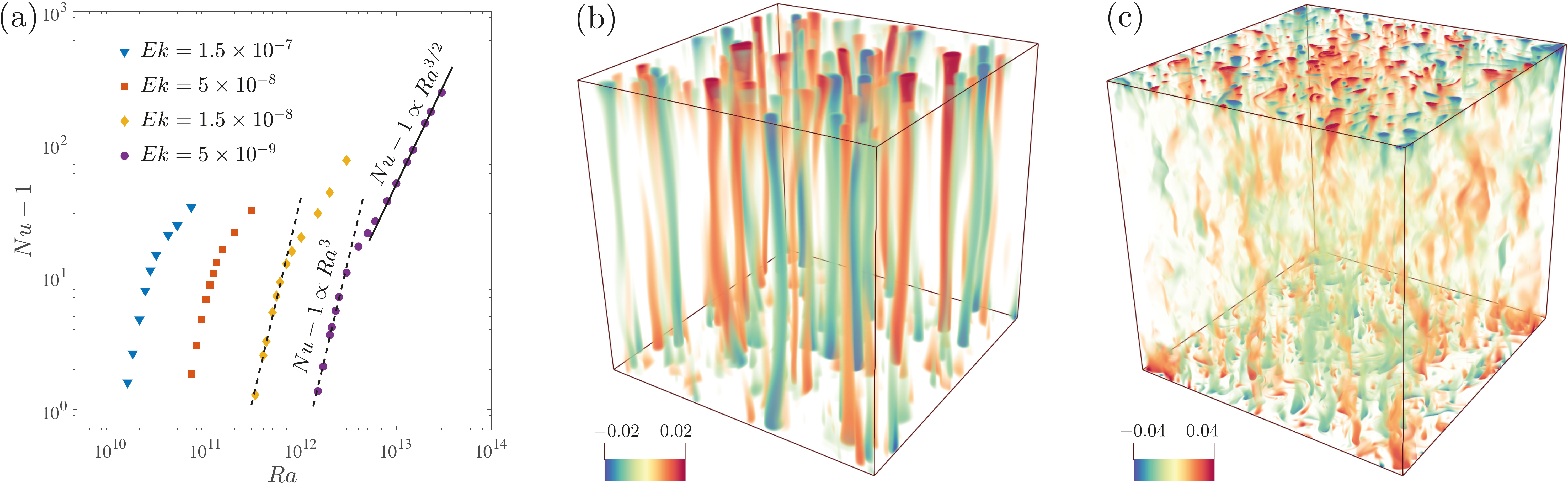}
\caption{
(a) Convective heat transport $Nu-1$ as a function $Ra$, for different $Ek$ and $Pr=1$.
All studied cases correspond to the rotation-dominated regime of RRBC (the Rossby number $Ro\ll1$). One can see that when $Ra$ is sufficiently large and $Ek$ sufficiently small, $Nu-1$ scales as $\propto Ra^3$ for relatively smaller values of $Ra$ and as $\propto Ra^{3/2}$ for larger $Ra$. 
The thermal fluctuations $(\theta - \langle \theta \rangle_A)/\Delta$ illustrate (b) the Taylor columns at $Ra=2.5\times10^{12}$ and (c) geostrophic turbulence at $Ra=10^{13}$ in these two regimes for $Ek=5\times10^{-9}$. Here, $\langle...\rangle_A$ denotes the average in time and over any horizontal cross-section $A$. For clarity, the domains are stretched horizontally by a factor of $8$ (see also Supplemental Material \cite{Supp}).
}
\label{fig:Space}
\end{figure*}

The scaling relations for the heat ($Nu$) and momentum ($Re$) transfer are usually sought as functions of $Ek$, $Ra$ and $Pr$, expressed in forms $\sim Ra^{\alpha}Ek^{\beta}Pr^{\gamma}$. 
For RRBC, under the assumption that the heat flux is independent of 
$\nu$ and $\kappa$, a diffusion-free heat transfer scaling law $Nu\sim Ra^{3/2}Ek^{2}Pr^{-1/2}$ can be derived \citep{Stevenson79, Gillet06, Julien12, Julien12a, Gastine16, Plumley17, Plumley19, Aurnou20, Bouillaut21}.
This relation is associated with the geostrophic turbulence regime, where not only the heat flux but also the whole system is independent of 
$\nu$ and $\kappa$
\citep{Schmitz09, Julien12, Plumley17, Guervilly19, Bouillaut21, Wang21}, following the Kolmogorov's energy cascade picture \citep{Ahlers09}. This regime has been also studied in a few experiments and DNS, mainly for free-slip boundary conditions (BCs) \citep{Julien12, Stellmach14, Plumley17, Bouillaut21}. There, to make the resolution in the DNS and rotation rate in experiments manageable, the typical $Ek$ lies in the order of $10^{-7}$ at least and $Ra$ in the order of $10^{12}$ at most.

In the limit of rapid rotation ($Ek\rightarrow0$), strong thermal forcing ($Ra\rightarrow\infty$) and infinite $Pr$, from the asymptotically reduced equations for $Ra\,Ek^{8/5}=\mathcal{O}(1)$ an upper bound $Nu\leq20.56\,(Ra/Ra_c)^3\propto \widetilde{Ra}^3$ was derived in \cite{Grooms2015}, where $Ra_c$ is the critical $Ra$ for the onset of RRBC and $\widetilde{Ra}\equiv Ra\,Ek^{4/3}$.
Here, $Nu$ increases much faster than in the regime of geostrophic turbulence.
One comes to a similar scaling relation, for any $Pr$, under the assumption that the total vertical heat flux is independent of the fluid layer depth $L$ (but not of $\kappa$ and $\nu$ as in the geostrophic turbulence regime).
This assumption immediately gives the scaling relations $Nu\propto Ra^{1/3}$ for the case of weak or no rotation \citep{Malkus1954, Priestley1959}, and $Nu\propto\widetilde{Ra}^3$ for the case of rotation dominance, see, e.g., \cite{Boubnov90, King12}. 
Although such scaling of $Nu$ with the control parameters was observed in some experiments and simulations for no-slip BCs at the plates and periodic lateral BCs \cite{King12, Stellmach14, Cheng15, Cheng16, Julien16, Lu21, Guzman21}, 
the behavior of $Re$ and typical length scales in that regime remain unclear.
Also unclear is how this regime is connected to the regime of geostrophic turbulence.

In this Letter, we present results of the DNS of RRBC for $Pr=1$ and extreme parameter range for 
$Ra$ from $1.5\times10^{10}$ to $3\times10^{13}$ and $Ek$ from $1.5\times10^{-7}$ down to $5\times10^{-9}$ (see Fig.~\ref{fig:Space}(a) for the parameter space).
The Navier--Stokes equations for the heat and momentum transport within the Boussinesq approximation
are solved numerically in a planar geometry with the no-slip BCs at the horizontal surfaces of a fluid layer, which is heated from below and cooled from above, subject to a constant rotation rate about a vertical axis (see Supplemental Material \cite{Supp}).
We achieve both geostrophic regimes in our DNS and show the scaling relations for $Nu$ and $Re$, the kinetic energy and thermal dissipation rates in both regimes, and offer a theoretical explanation for them.
The transition between the two regimes is seen in the scalings of all quantities, however, the scaling with $Ra$ and $Ek$ of the convective bulk length scale $\ell$ remains the same in both regimes.
Note that this transition between the two rotation-dominated regimes is of course very different from the transition between the rotation dominance to the gravitational buoyancy dominance in RRBC \cite{Vorobieff2002, Kunnen2008, King09, Stevens09, Stevens10, Weiss2011, Weiss2011b, King13, Ecke14, Horn14, Stellmach14, Horn2015, Cheng15, Cheng18, Cheng20, Kunnen21, Hartmann22, Ecke23}. 
For a discussion of the latter transition we refer to \cite[][\S 3.3]{Ecke23}.

\begin{figure*}
\centering\includegraphics[width=1\linewidth]{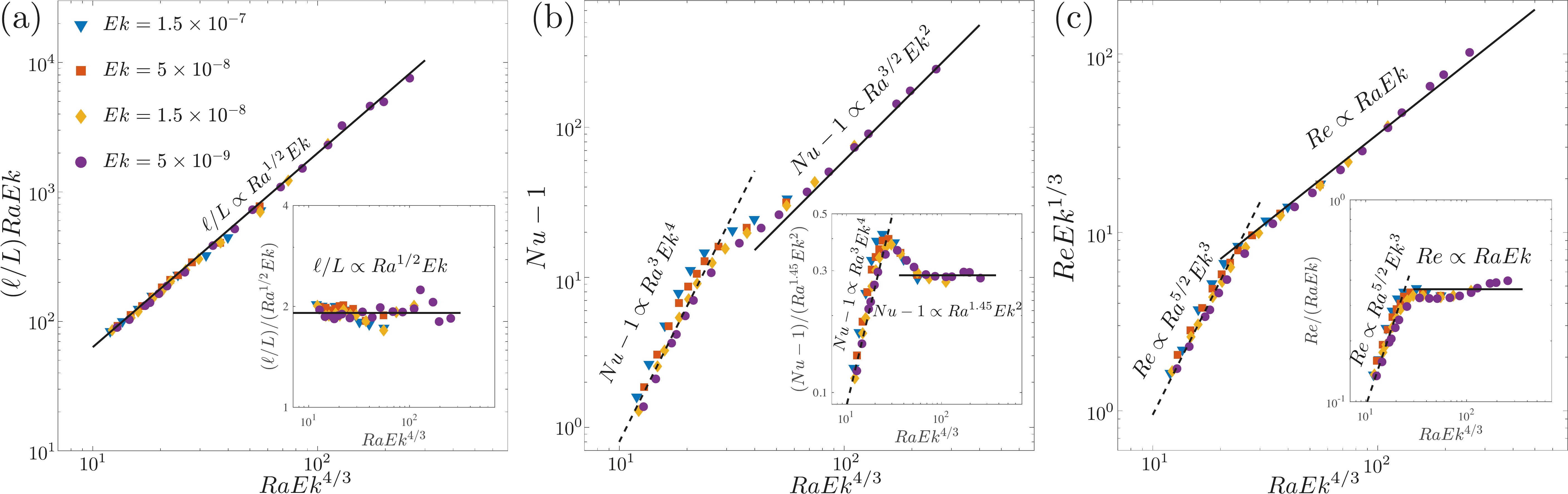}
\caption{
Dimensionless convective (a) length scale $\ell/L$ (multiplied by $RaEk$), (b) heat transport $Nu-1$ and (c) $Re$ (multiplied by $Ek^{1/3}$) as functions of $\widetilde{Ra}$, for all DNS data from Fig.~\ref{fig:Space}(a). (a) The DNS demonstrate $\ell/L\sim Ro$. Here, $\ell/L$ is evaluated using 2D Fourier transforms of the vertical velocity $u_z$ (see the main text for details). The inset shows $\ell/L$, normalized by $Ro$. (b) The data for different $Ek$ fall on one graph. For larger $\widetilde{Ra}$, $Nu-1$ scales as expected for the geostrophic turbulence regime: $Nu-1 \propto \widetilde{Ra}^{3/2}$ (solid line), while for smaller $\widetilde{Ra}$ one observes a transition to a regime with $Nu-1 \propto \widetilde{Ra}^{3}$ (dashed line). In the inset plot, the same data are presented in a compensated way, where the normalization is chosen according to the best fit of the data for $Ek=5\times10^{-9}$ and large values of $Ra$, $Nu-1\propto Ra^{1.45\pm0.07}$. 
(c) For larger $\widetilde{Ra}$, $Re$ scales almost as expected for the geostrophic turbulence regime: $Re \propto Ra\,Ek$ (solid line), while for smaller $\widetilde{Ra}$, it scales as $Re \propto Ra^{5/2}Ek^3$ (dashed line). In the inset plot, $Re$, normalized by its scaling in the geostrophic diffusion-free regime, i.e., $Re/(RaEk)$, is plotted versus $\widetilde{Ra}$.
}
\label{fig:LeNuRe}
\end{figure*}

In what follows we assume that in any rotation-dominated regime the dimensionless convective heat transport is proportional to a power function of the supercriticality parameter $\widetilde{Ra}\equiv Ra\,Ek^{4/3}$ 
\cite[see, e.g.,][]{Julien12, Stellmach14},
\begin{equation}
Nu-1\propto \widetilde{Ra}^\xi,
 \label{1}
\end{equation}
with different exponents $\xi$ in different regimes.

First we discuss relations that hold in both studied rotation-dominated regimes and recall the rigorous relations for the time- and volume-averaged kinetic energy dissipation rate $\epsilon_u=\langle \nu(\partial_iu_j(\boldsymbol{x},t))^2 \rangle$ and thermal dissipation rate $\epsilon_\theta=\langle \kappa(\partial_i\theta(\boldsymbol{x},t))^2 \rangle$
that hold in RRBC \citep{Ahlers09}:
\begin{eqnarray}
  \epsilon_u&=& ({\nu^3}/{L^4})(Nu-1)RaPr^{-2}, \label{3}\\
  \epsilon_\theta&=& \kappa({\Delta^2}/{L^2})Nu.
\label{4}
\end{eqnarray}
We introduce $u$, $\theta$ and $\ell$, which are the representative convective scales for, respectively, the velocity, temperature and length.
The total heat flux can be decomposed into a conductive contribution $\kappa\Delta/L$ and a convective contribution $q$ that scales as $q\sim u\theta$.
The dimensionless convective heat flux scales then as
\begin{eqnarray}
  Nu-1\sim \frac{q}{\kappa\Delta/L}\sim \frac{u\theta}{\kappa\Delta/L}.
\label{5}
\end{eqnarray}
Analogously, the total thermal dissipation rate $\epsilon_\theta$ can be decomposed into a conductive contribution $\kappa{\Delta^2}/{L^2}$ and a convective contribution $\widetilde \epsilon_{\theta}$, which scales as $\widetilde \epsilon_{\theta}\sim{u\theta^2}/{\ell}$.
This, in combination with Eq.~(\ref{4}), gives
\begin{eqnarray}
  Nu-1\sim \frac{\widetilde \epsilon_{\theta}}{\kappa\,\Delta^2/L^2}\sim\frac{\theta^2}{\Delta^2}\frac{L}{\ell}\frac{uL}{\nu}\frac{\nu}{\kappa}.
\label{6}
\end{eqnarray}
Combining Eq.~(\ref{5}) and Eq.~(\ref{6}) we obtain
$\ell/L\sim\theta/\Delta$,
which together with Eq.~(\ref{6}) leads to 
\begin{eqnarray}
  Nu-1\sim \frac{\theta^2}{\Delta^2}\frac{L}{\ell}\,Re\,Pr\sim \frac{\ell}{L}\,Re\,Pr.
\label{7}
\end{eqnarray}

In a turbulent flow, $\epsilon_u$ scales as $\epsilon_u \sim {u^3}/{\ell}$ \cite{Landau1987}.
This, in combination with Eq.~(\ref{3}) and Eq.~(\ref{7}) leads to 
\begin{eqnarray}
 Re&\sim& ({\ell}/{L})\,Pr^{-1/2}Ra^{1/2}, \label{8}\\
 Nu-1&\sim& ({\ell}/{L})^2\,Pr^{1/2}Ra^{1/2}. \label{9}
\end{eqnarray}
The dimensional convective bulk length scale $\ell$ is diffusion-free in the geostrophic turbulence regime, meaning that it is independent of $\nu$ and $\kappa$.
If $\ell/L$ is thought as a product of power functions of $Ra$, $Ek$ and $Pr$, then the requirement for $\ell/L$ to be diffusion-free, i.e. $\ell/L\propto \nu^0\kappa^0$, means that $\ell/L$ must scale as
\begin{equation}
\ell/L\sim Ra^a\,Ek^{2a}\,Pr^{-a}
 \label{2}
\end{equation}
for some value of $a$.
From Eqs.~(\ref{9}) and (\ref{2}) it follows that 
\begin{eqnarray}
Nu-1\propto \widetilde{Ra}^{2a+1/2}Ek^{(4a-2)/3}. \label{10}
\end{eqnarray}
From Eqs.~(\ref{1}) and (\ref{10}) we derive $a=1/2$ and $\xi=3/2$, 
and, therefore, the following relations must be fulfilled:
\begin{eqnarray}
      \ell/L &\sim& Ra^{1/2}\,Ek\,Pr^{-1/2},\label{11}\\
      Nu-1 &\sim& Ra^{3/2}\,Ek^{2}\,Pr^{-1/2},\label{13}\\
      Re &\sim& Ra\,Ek\,Pr^{-1},\label{12}\\
      (L^4/\nu^3)\epsilon_u&\sim& Ra^{5/2}\,Ek^{2}\,Pr^{-5/2}.\label{13a}
\end{eqnarray}
Note that $a=1/2$ means that $\ell/L$ scales as the Rossby number $Ro\equiv \sqrt{Ra/Pr}Ek$.
Eqs.~(\ref{11})--(\ref{13a}) show that in the geostrophic turbulence regime, the dimensional convective bulk length scale $\ell$ and velocity scale $\nu Re/L$, the convective heat flux $\kappa\Delta/L(Nu-1)$ and the dissipation rates are all diffusion-free; they all scale as $\propto\nu^0\kappa^0$.
The scaling relations Eqs.~(\ref{11})--(\ref{12}) for the geostrophic turbulence were proposed in \cite{Aurnou20}, see also \citep{Ecke23, Sprague2006, Julien12a, Plumley16, Plumley17, Plumley19, Guervilly19, Guzman21}.

One can argue that $\ell$ can be non-dimensionalized (without involving $\nu$ or $\kappa$) not only with $L$ but in another way, using, e.g., $\alpha_T\Delta g/\Omega^2$ as the reference length.
In that case the diffusion-free length scale would imply 
$\ell/(\alpha_T\Delta g/\Omega^2)\sim Ra^b\,Ek^{2b}\,Pr^{-b}$ 
for some $b$, which is equivalent to 
$\ell/L\sim Ra^{1+b}\,Ek^{2b+2}\,Pr^{-1-b}$.
Combining this with Eqs.~(\ref{9}) and (\ref{1}), we derive that $b=-1/2$ and $\xi=3/2$, and that the scaling relations for the geostrophic turbulence, Eqs.~(\ref{11})--(\ref{13a}), should anyway hold.

To verify the scaling relations~(\ref{11})--(\ref{13a}), we have conducted DNS of RRBC in domains with periodic lateral BCs, in order to avoid the effect of the wall modes \cite{Rossby1969, Ecke1992, Goldstein1993, Favier2020, Shishkina2020} or boundary zonal flows \cite{Zhang20, Wit2020, Zhang21, Ecke22, Wedi22}.
The studied DNS parameter range is unprecedented: $Ra$ up to $3\times 10^{13}$ and $Ek$ down to $ 5\times 10^{-9}$.

First we verify that $\ell/L$ scales according to Eq.~(\ref{11}).
It is indeed fulfilled, since $(\ell/L)Ra\,Ek$  scales as 
\begin{eqnarray}
   (\ell/L)Ra\,Ek \propto Ra^{3/2}\,Ek^2= \widetilde{Ra}^{3/2}.\label{20}
\end{eqnarray}
This is supported by the DNS data presented in Fig.~\ref{fig:LeNuRe}(a).
Here, following \cite{Guervilly14, Guervilly19, Maffei21}, we conduct the 2D Fourier transforms of the instantaneous vertical velocity $u_z$ at the mid height, in order to evaluate $\ell/L$ as follows:
$\ell/L=\sum_{k_h} [\hat{u}_z(k_h)\hat{u}^*_z(k_h)]/\sum_{k_h} k_h [\hat{u}_z(k_h)\hat{u}^*_z(k_h)]$, where $\hat{u}_z(k_h)$ and $\hat{u}^*_z(k_h)$ are, respectively, the 2D Fourier transforms of $u_z$ and its 
complex conjugate
and $k_h\equiv(k_x^2+k_y^2)^{1/2}$ is the horizontal wavenumber.
The usage of other quantities to evaluate the convective length scale leads to similar results (see Supplemental Material \cite{Supp}).

\begin{figure*}[t]\centering
\includegraphics[width=1\linewidth]{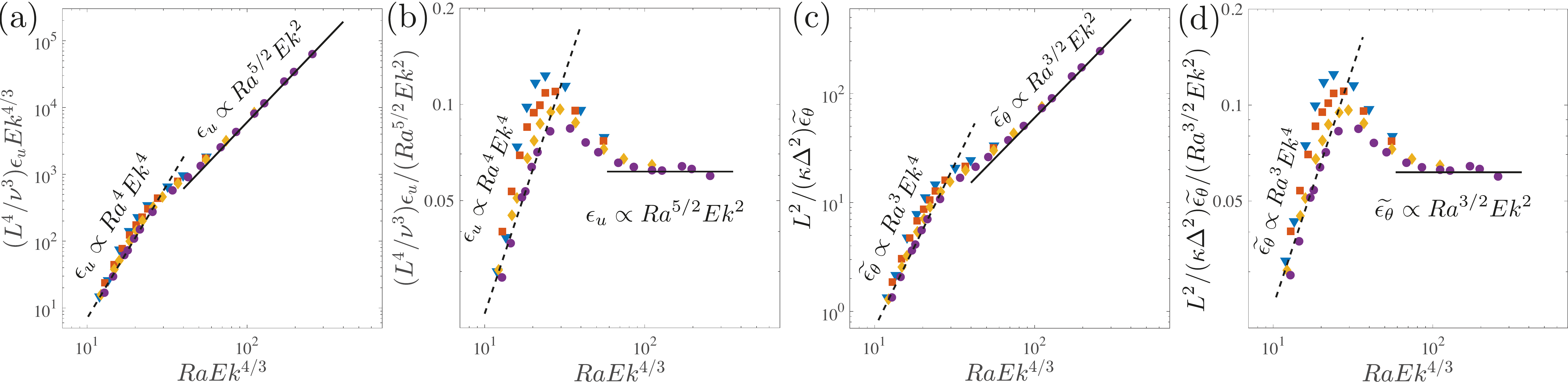}
\caption{
The dependence on $\widetilde{Ra}$ of the time- and volume-averaged (a-b) kinetic dissipation rate $\epsilon_u$ and (c-d) convective thermal dissipation rate $\widetilde \epsilon_{\theta}\equiv\epsilon_{\theta}-\kappa\Delta^2/L^2$.
(a) For larger $\widetilde{Ra}$, $\epsilon_u$ scales as expected for the geostrophic turbulence regime: $\epsilon_u \propto Ra^{5/2}Ek^2$ (solid line), while for smaller $\widetilde{Ra}$, it scales as $\epsilon_u \propto Ra^{4}Ek^4$ (dashed line).
(b) $\epsilon_u$, normalized by $Ra^{5/2}Ek^2$. 
(c) For larger $\widetilde{Ra}$, $\widetilde \epsilon_{\theta}$ scales as expected for the geostrophic turbulence regime: $\widetilde \epsilon_{\theta} \propto Ra^{3/2}Ek^2$ (solid line), while for smaller $\widetilde{Ra}$, it scales as $\widetilde \epsilon_{\theta}\propto Ra^{3}Ek^4$ (dashed line). 
(d) $\widetilde \epsilon_{\theta}$, normalized by $Ra^{3/2}Ek^2$.
Symbols have the same meaning as in Figs.~\ref{fig:Space}(a) and \ref{fig:LeNuRe}.
}
\label{fig:EuEt}
\end{figure*}

As it was assumed in Eq.~(\ref{1}), $Nu-1$ behaves indeed as a function of $\widetilde{Ra}$, since all data from Fig.~\ref{fig:Space}(a) follow a master curve when plotted versus $\widetilde{Ra}$, see Fig.~\ref{fig:LeNuRe}(b).
For large values of $\widetilde{Ra}$, $Nu-1$ scales according to Eq.~(\ref{13}), as expected.
At $\widetilde{Ra}$ about 30, one observes a transition to some other regime, with a steeper growth of $Nu$.

To verify the theoretical predictions on the $Re$-scaling in the geostrophic turbulence regime, 
we notice that $Re\,Ek^{1/3}$ should scale as $\propto \widetilde{Ra}$ if Eq.~(\ref{12}) is fulfilled.
Indeed, the data in Fig.~\ref{fig:LeNuRe}(c) support this scaling relation for large $\widetilde{Ra}$.
Here, following \cite{Guervilly14, Gastine16, Guervilly19, Maffei21}, in order to minimize the impact from the large-scale vortices (LSVs) and properly characterize the amplitude of convective bulk motions and evaluate $Re$, we use the vertical velocity $u_z$ as the typical velocity scale: $Re=\sqrt{\langle u_z^2\rangle}L/\nu$.
Note that our DNS as well as previous DNS for periodic lateral BCs show formation of LSVs in the geostrophic turbulence regime, which is also associated with an additional increase of $Re$ for larger $\widetilde{Ra}$ \citep{Julien12, Julien12a, Guervilly14, Guervilly17, Julien18, Favier19, Wit22}.

Finally we verify the scaling relations for the dissipation rates.
In Fig.~\ref{fig:EuEt}(a-b) we observe the scaling $(L^4/\nu^3)\epsilon_u\,Ek^{4/3}\propto \widetilde{Ra}^{5/2}$ for large $\widetilde{Ra}$, meaning that 
$\epsilon_u$ follows Eq.~(\ref{13a}) in the geostrophic turbulence regime.
The convective thermal dissipation rate, $\widetilde \epsilon_{\theta}$, behaves similarly to $Nu-1$ and the data for $\widetilde \epsilon_{\theta}$ also follow the expected scaling relations, see Fig.~\ref{fig:EuEt}(c-d).

Thus all shown scalings of $\ell/L$, $Nu-1$, $Re$, $\epsilon_u$ and $\widetilde \epsilon_{\theta}$ for large $\widetilde{Ra}$ follow the predictions (\ref{11})--(\ref{13a}) for the geostrophic turbulence regime.
But what is the regime of a steeper growth of $Nu-1$, $Re$, $\epsilon_u$ and $\widetilde \epsilon_{\theta}$ that we observe for smaller $\widetilde{Ra}$ ($\widetilde{Ra}\lesssim30$) in Fig.~\ref{fig:LeNuRe} and Fig.~\ref{fig:EuEt}?
How can we understand its scaling relations theoretically?

For any given $Ek$, with decreasing $\widetilde{Ra}$, the flow should gradually laminarize, and the $\epsilon_u$-scaling should undergo transition to the scaling $\epsilon_u \sim \nu {u^2}/{\ell^2}$.
This relation in combination with Eq.~(\ref{3}) and Eq.~(\ref{7}) leads to
\begin{eqnarray}
 Re&\sim& ({\ell}/{L})^3\,Pr^{-1}Ra, \label{14}\\
 Nu-1&\sim& ({\ell}/{L})^4\,Ra. \label{15}
\end{eqnarray}
The DNS data show that in this regime of lower $\widetilde{Ra}$, the convective bulk length scale $\ell/L$ is also proportional to $Ra^{1/2}\,Ek$ (with possible unknown $Pr$ dependence), see Fig.~\ref{fig:LeNuRe}(a).
From this, Eq.~(\ref{15}) and Eq.~(\ref{1}) we obtain $\xi=3$, meaning a steeper growth of $Re$ and $Nu$ with increasing $\widetilde{Ra}$:
\begin{eqnarray}
      Nu-1 &\propto& Ra^{3}\,Ek^{4},\label{18}\\
      Re &\propto& Ra^{5/2}\,Ek^3,\label{17}\\
     (L^4/\nu^3)\epsilon_u&\propto&Ra^{4}\,Ek^{4}.\label{18a}
\end{eqnarray}
The steep $Nu$-scaling, Eq.~(\ref{18}), was derived in \cite{Boubnov90}, where the marginal thermal boundary layer instability in rotating convection was considered, see also \cite{King12, Stellmach14, Cheng15, Cheng16, Julien16, Guzman21}.

Thus we propose that with decreasing $\widetilde{Ra}$, the regime of geostrophic turbulence, Eq.~(\ref{11})--(\ref{13a}), should undergo transition to another scaling regime, Eq.~(\ref{18})--(\ref{18a}), and in both regimes $\ell/L \propto Ra^{1/2}\,Ek$ hold. 
And indeed, our DNS data fully support this.
For smaller $\widetilde{Ra}$, $Nu-1$ follows the relation (\ref{18}), while for larger $\widetilde{Ra}$ it scales according to Eq.~(\ref{13}), see Fig.~\ref{fig:LeNuRe}(b).
The theory says that $Re\,Ek^{1/3}$ should scale as $\propto \widetilde{Ra}$ in the geostrophic turbulence regime, Eq.~(\ref{12}), and as $\propto \widetilde{Ra}^{5/2}$ in another regime, Eq.~(\ref{17}), and indeed, the data in Fig.~\ref{fig:LeNuRe}(c) support these scaling relations.
With the decreasing $\widetilde{Ra}$, the scaling of $\epsilon_u$ changes from 
Eq.~(\ref{13a}) to Eq.~(\ref{18a}), meaning that the scaling of $(L^4/\nu^3)\epsilon_u\,Ek^{4/3}$ changes from 
$\propto \widetilde{Ra}^{5/2}$ to $\propto \widetilde{Ra}^{4}$, and the data in  Fig.~\ref{fig:EuEt}(a-b) clearly support this.
\color{black}

To sum up, based on our DNS of RRBC for extreme $Ra$ and $Ek$ we have verified the existence and the scaling relations in the geostrophic turbulence regime, Eq.~(\ref{11})--(\ref{13a}).
We furthermore have shown that this regime is connected to another rotation-dominated regime, Eq.~(\ref{18})--(\ref{18a}), which can be achieved by decreasing the thermal driving ($Ra$), while keeping constant rotation ($Ek$).
In both cases the convective heat transport, $Nu-1$, scales as a power function of $\widetilde{Ra}$, with a power of 3/2 in the geostrophic turbulence regime and a power of 3 in the regime of the steep growth.
The principle difference between the two regimes is the different scaling of the kinetic energy dissipation rate: it is turbulent in one case and laminar in the other.

For lab experiments and DNS, a more extreme $Ra$ and $Ek$ range, for different $Pr$, is desired, in order to study in detail the geostrophic turbulence regime and its transition to the buoyancy-dominated regime.

\begin{acknowledgments}
We thank J.M.~Aurnou, R.E.~Ecke, K.~Julien and D.~Lohse for illuminating discussions and gratefully acknowledge the support from the Max Planck Society, the Alexander von Humboldt Foundation and German Research Foundation (DFG), and the computing time provided 
on the high-performance computer Lichtenberg at the NHR Centers NHR4CES,
the HPC systems of Max Planck Computing and Data Facility (MPCDF), 
the HoreKa supercomputer,
and the GCS Supercomputer SuperMUC-NG at Leibniz Supercomputing Centre.

\end{acknowledgments}

\bibliography{bib}

\begin{thebibliography}{77}%
\makeatletter
\providecommand \@ifxundefined [1]{%
 \@ifx{#1\undefined}
}%
\providecommand \@ifnum [1]{%
 \ifnum #1\expandafter \@firstoftwo
 \else \expandafter \@secondoftwo
 \fi
}%
\providecommand \@ifx [1]{%
 \ifx #1\expandafter \@firstoftwo
 \else \expandafter \@secondoftwo
 \fi
}%
\providecommand \natexlab [1]{#1}%
\providecommand \enquote  [1]{``#1''}%
\providecommand \bibnamefont  [1]{#1}%
\providecommand \bibfnamefont [1]{#1}%
\providecommand \citenamefont [1]{#1}%
\providecommand \href@noop [0]{\@secondoftwo}%
\providecommand \href [0]{\begingroup \@sanitize@url \@href}%
\providecommand \@href[1]{\@@startlink{#1}\@@href}%
\providecommand \@@href[1]{\endgroup#1\@@endlink}%
\providecommand \@sanitize@url [0]{\catcode `\\12\catcode `\$12\catcode
  `\&12\catcode `\#12\catcode `\^12\catcode `\_12\catcode `\%12\relax}%
\providecommand \@@startlink[1]{}%
\providecommand \@@endlink[0]{}%
\providecommand \url  [0]{\begingroup\@sanitize@url \@url }%
\providecommand \@url [1]{\endgroup\@href {#1}{\urlprefix }}%
\providecommand \urlprefix  [0]{URL }%
\providecommand \Eprint [0]{\href }%
\providecommand \doibase [0]{http://dx.doi.org/}%
\providecommand \selectlanguage [0]{\@gobble}%
\providecommand \bibinfo  [0]{\@secondoftwo}%
\providecommand \bibfield  [0]{\@secondoftwo}%
\providecommand \translation [1]{[#1]}%
\providecommand \BibitemOpen [0]{}%
\providecommand \bibitemStop [0]{}%
\providecommand \bibitemNoStop [0]{.\EOS\space}%
\providecommand \EOS [0]{\spacefactor3000\relax}%
\providecommand \BibitemShut  [1]{\csname bibitem#1\endcsname}%
\let\auto@bib@innerbib\@empty
\bibitem [{\citenamefont {Ecke}\ and\ \citenamefont
  {Shishkina}(2023)}]{Ecke23}%
  \BibitemOpen
  \bibfield  {author} {\bibinfo {author} {\bibfnamefont {R.~E.}\ \bibnamefont
  {Ecke}}\ and\ \bibinfo {author} {\bibfnamefont {O.}~\bibnamefont
  {Shishkina}},\ }\href@noop {} {\bibfield  {journal} {\bibinfo  {journal}
  {Annu. Rev. Fluid Mech.}\ }\textbf {\bibinfo {volume} {55}},\ \bibinfo
  {pages} {603–638} (\bibinfo {year} {2023})}\BibitemShut {NoStop}%
\bibitem [{\citenamefont {Kunnen}(2021)}]{Kunnen21}%
  \BibitemOpen
  \bibfield  {author} {\bibinfo {author} {\bibfnamefont {R.~P.~J.}\
  \bibnamefont {Kunnen}},\ }\href@noop {} {\bibfield  {journal} {\bibinfo
  {journal} {J. Turbulence}\ }\textbf {\bibinfo {volume} {22}},\ \bibinfo
  {pages} {267–296} (\bibinfo {year} {2021})}\BibitemShut {NoStop}%
\bibitem [{\citenamefont {Busse}\ and\ \citenamefont
  {Carrigan}(1976)}]{Busse76}%
  \BibitemOpen
  \bibfield  {author} {\bibinfo {author} {\bibfnamefont {F.~H.}\ \bibnamefont
  {Busse}}\ and\ \bibinfo {author} {\bibfnamefont {C.~R.}\ \bibnamefont
  {Carrigan}},\ }\href@noop {} {\bibfield  {journal} {\bibinfo  {journal}
  {Science}\ }\textbf {\bibinfo {volume} {191}},\ \bibinfo {pages} {81}
  (\bibinfo {year} {1976})}\BibitemShut {NoStop}%
\bibitem [{\citenamefont {Heimpel}\ \emph {et~al.}(2005)\citenamefont
  {Heimpel}, \citenamefont {Gastine},\ and\ \citenamefont {Wicht}}]{Heimpel05}%
  \BibitemOpen
  \bibfield  {author} {\bibinfo {author} {\bibfnamefont {M.}~\bibnamefont
  {Heimpel}}, \bibinfo {author} {\bibfnamefont {T.}~\bibnamefont {Gastine}}, \
  and\ \bibinfo {author} {\bibfnamefont {J.}~\bibnamefont {Wicht}},\
  }\href@noop {} {\bibfield  {journal} {\bibinfo  {journal} {Nature}\ }\textbf
  {\bibinfo {volume} {438}},\ \bibinfo {pages} {193–196} (\bibinfo {year}
  {2005})}\BibitemShut {NoStop}%
\bibitem [{\citenamefont {Ahlers}\ \emph {et~al.}(2009)\citenamefont {Ahlers},
  \citenamefont {Grossmann},\ and\ \citenamefont {Lohse}}]{Ahlers09}%
  \BibitemOpen
  \bibfield  {author} {\bibinfo {author} {\bibfnamefont {G.}~\bibnamefont
  {Ahlers}}, \bibinfo {author} {\bibfnamefont {S.}~\bibnamefont {Grossmann}}, \
  and\ \bibinfo {author} {\bibfnamefont {D.}~\bibnamefont {Lohse}},\
  }\href@noop {} {\bibfield  {journal} {\bibinfo  {journal} {Rev. Mod. Phys.}\
  }\textbf {\bibinfo {volume} {81}},\ \bibinfo {pages} {503} (\bibinfo {year}
  {2009})}\BibitemShut {NoStop}%
\bibitem [{\citenamefont {Wicht}\ and\ \citenamefont
  {Sanchez}(2019)}]{Wicht19}%
  \BibitemOpen
  \bibfield  {author} {\bibinfo {author} {\bibfnamefont {J.}~\bibnamefont
  {Wicht}}\ and\ \bibinfo {author} {\bibfnamefont {S.}~\bibnamefont
  {Sanchez}},\ }\href@noop {} {\bibfield  {journal} {\bibinfo  {journal}
  {Geophys. Astrophys. Fluid Dyn.}\ }\textbf {\bibinfo {volume} {113}},\
  \bibinfo {pages} {1} (\bibinfo {year} {2019})}\BibitemShut {NoStop}%
\bibitem [{\citenamefont {Yadav}\ and\ \citenamefont
  {Bloxham}(2020)}]{Yadav20a}%
  \BibitemOpen
  \bibfield  {author} {\bibinfo {author} {\bibfnamefont {R.~K.}\ \bibnamefont
  {Yadav}}\ and\ \bibinfo {author} {\bibfnamefont {J.}~\bibnamefont
  {Bloxham}},\ }\href@noop {} {\bibfield  {journal} {\bibinfo  {journal} {Proc.
  Natl. Acad. Sci. U.S.A.}\ }\textbf {\bibinfo {volume} {117}},\ \bibinfo
  {pages} {13991–13996} (\bibinfo {year} {2020})}\BibitemShut {NoStop}%
\bibitem [{\citenamefont {Yadav}\ \emph {et~al.}(2020)\citenamefont {Yadav},
  \citenamefont {Heimpel},\ and\ \citenamefont {Bloxham}}]{Yadav20b}%
  \BibitemOpen
  \bibfield  {author} {\bibinfo {author} {\bibfnamefont {R.~K.}\ \bibnamefont
  {Yadav}}, \bibinfo {author} {\bibfnamefont {M.}~\bibnamefont {Heimpel}}, \
  and\ \bibinfo {author} {\bibfnamefont {J.}~\bibnamefont {Bloxham}},\
  }\href@noop {} {\bibfield  {journal} {\bibinfo  {journal} {Sci. Adv.}\
  }\textbf {\bibinfo {volume} {6}},\ \bibinfo {pages} {eabb9298} (\bibinfo
  {year} {2020})}\BibitemShut {NoStop}%
\bibitem [{\citenamefont {Jones}(2011)}]{Jones11}%
  \BibitemOpen
  \bibfield  {author} {\bibinfo {author} {\bibfnamefont {C.~A.}\ \bibnamefont
  {Jones}},\ }\href@noop {} {\bibfield  {journal} {\bibinfo  {journal} {Annu.
  Rev. Fluid Mech.}\ }\textbf {\bibinfo {volume} {43}},\ \bibinfo {pages} {583}
  (\bibinfo {year} {2011})}\BibitemShut {NoStop}%
\bibitem [{\citenamefont {Aurnou}\ \emph {et~al.}(2015)\citenamefont {Aurnou},
  \citenamefont {Calkins}, \citenamefont {Cheng}, \citenamefont {Julien},
  \citenamefont {King}, \citenamefont {Nieves}, \citenamefont {Soderlund},\
  and\ \citenamefont {Stellmach}}]{Aurnou15}%
  \BibitemOpen
  \bibfield  {author} {\bibinfo {author} {\bibfnamefont {J.~M.}\ \bibnamefont
  {Aurnou}}, \bibinfo {author} {\bibfnamefont {M.~A.}\ \bibnamefont {Calkins}},
  \bibinfo {author} {\bibfnamefont {J.~S.}\ \bibnamefont {Cheng}}, \bibinfo
  {author} {\bibfnamefont {K.}~\bibnamefont {Julien}}, \bibinfo {author}
  {\bibfnamefont {E.}~\bibnamefont {King}}, \bibinfo {author} {\bibfnamefont
  {D.}~\bibnamefont {Nieves}}, \bibinfo {author} {\bibfnamefont {K.~M.}\
  \bibnamefont {Soderlund}}, \ and\ \bibinfo {author} {\bibfnamefont
  {S.}~\bibnamefont {Stellmach}},\ }\href@noop {} {\bibfield  {journal}
  {\bibinfo  {journal} {Phys. Earth Planet. Inter.}\ }\textbf {\bibinfo
  {volume} {246}},\ \bibinfo {pages} {52–71} (\bibinfo {year}
  {2015})}\BibitemShut {NoStop}%
\bibitem [{\citenamefont {Cheng}\ \emph {et~al.}(2015)\citenamefont {Cheng},
  \citenamefont {Stellmach}, \citenamefont {Ribeiro}, \citenamefont {Grannan},
  \citenamefont {King},\ and\ \citenamefont {Aurnou}}]{Cheng15}%
  \BibitemOpen
  \bibfield  {author} {\bibinfo {author} {\bibfnamefont {J.~S.}\ \bibnamefont
  {Cheng}}, \bibinfo {author} {\bibfnamefont {S.}~\bibnamefont {Stellmach}},
  \bibinfo {author} {\bibfnamefont {A.}~\bibnamefont {Ribeiro}}, \bibinfo
  {author} {\bibfnamefont {A.}~\bibnamefont {Grannan}}, \bibinfo {author}
  {\bibfnamefont {E.~M.}\ \bibnamefont {King}}, \ and\ \bibinfo {author}
  {\bibfnamefont {J.~M.}\ \bibnamefont {Aurnou}},\ }\href@noop {} {\bibfield
  {journal} {\bibinfo  {journal} {Geophys. J. Int.}\ }\textbf {\bibinfo
  {volume} {201}},\ \bibinfo {pages} {1} (\bibinfo {year} {2015})}\BibitemShut
  {NoStop}%
\bibitem [{\citenamefont {Hanasoge}\ \emph {et~al.}(2016)\citenamefont
  {Hanasoge}, \citenamefont {Gizon},\ and\ \citenamefont
  {Sreenivasan}}]{Hanasoge16}%
  \BibitemOpen
  \bibfield  {author} {\bibinfo {author} {\bibfnamefont {S.}~\bibnamefont
  {Hanasoge}}, \bibinfo {author} {\bibfnamefont {L.}~\bibnamefont {Gizon}}, \
  and\ \bibinfo {author} {\bibfnamefont {K.~R.}\ \bibnamefont {Sreenivasan}},\
  }\href@noop {} {\bibfield  {journal} {\bibinfo  {journal} {Annu. Rev. Fluid
  Mech.}\ }\textbf {\bibinfo {volume} {48}},\ \bibinfo {pages} {191} (\bibinfo
  {year} {2016})}\BibitemShut {NoStop}%
\bibitem [{\citenamefont {Yadav}\ \emph {et~al.}(2016)\citenamefont {Yadav},
  \citenamefont {Gastine}, \citenamefont {Christensen}, \citenamefont
  {Duarte},\ and\ \citenamefont {Reiners}}]{Yadav16}%
  \BibitemOpen
  \bibfield  {author} {\bibinfo {author} {\bibfnamefont {R.~K.}\ \bibnamefont
  {Yadav}}, \bibinfo {author} {\bibfnamefont {T.}~\bibnamefont {Gastine}},
  \bibinfo {author} {\bibfnamefont {U.~R.}\ \bibnamefont {Christensen}},
  \bibinfo {author} {\bibfnamefont {L.~D.~V.}\ \bibnamefont {Duarte}}, \ and\
  \bibinfo {author} {\bibfnamefont {A.}~\bibnamefont {Reiners}},\ }\href@noop
  {} {\bibfield  {journal} {\bibinfo  {journal} {Geophys. J. Int.}\ }\textbf
  {\bibinfo {volume} {204}},\ \bibinfo {pages} {1120–1133} (\bibinfo {year}
  {2016})}\BibitemShut {NoStop}%
\bibitem [{\citenamefont {Cheng}\ \emph {et~al.}(2018)\citenamefont {Cheng},
  \citenamefont {Aurnou}, \citenamefont {Julien},\ and\ \citenamefont
  {Kunnen}}]{Cheng18}%
  \BibitemOpen
  \bibfield  {author} {\bibinfo {author} {\bibfnamefont {J.~S.}\ \bibnamefont
  {Cheng}}, \bibinfo {author} {\bibfnamefont {J.~M.}\ \bibnamefont {Aurnou}},
  \bibinfo {author} {\bibfnamefont {K.}~\bibnamefont {Julien}}, \ and\ \bibinfo
  {author} {\bibfnamefont {R.~P.~J.}\ \bibnamefont {Kunnen}},\ }\href@noop {}
  {\bibfield  {journal} {\bibinfo  {journal} {Geophys. Astrophys. Fluid Dyn.}\
  }\textbf {\bibinfo {volume} {112}},\ \bibinfo {pages} {277} (\bibinfo {year}
  {2018})}\BibitemShut {NoStop}%
\bibitem [{\citenamefont {Guervilly}\ \emph {et~al.}(2019)\citenamefont
  {Guervilly}, \citenamefont {Cardin},\ and\ \citenamefont
  {Schaeffer}}]{Guervilly19}%
  \BibitemOpen
  \bibfield  {author} {\bibinfo {author} {\bibfnamefont {C.}~\bibnamefont
  {Guervilly}}, \bibinfo {author} {\bibfnamefont {P.}~\bibnamefont {Cardin}}, \
  and\ \bibinfo {author} {\bibfnamefont {N.}~\bibnamefont {Schaeffer}},\
  }\href@noop {} {\bibfield  {journal} {\bibinfo  {journal} {Nature}\ }\textbf
  {\bibinfo {volume} {570}},\ \bibinfo {pages} {368} (\bibinfo {year}
  {2019})}\BibitemShut {NoStop}%
\bibitem [{\citenamefont {Schumacher}\ and\ \citenamefont
  {Sreenivasan}(2020)}]{Schumacher20}%
  \BibitemOpen
  \bibfield  {author} {\bibinfo {author} {\bibfnamefont {J.}~\bibnamefont
  {Schumacher}}\ and\ \bibinfo {author} {\bibfnamefont {K.~R.}\ \bibnamefont
  {Sreenivasan}},\ }\href@noop {} {\bibfield  {journal} {\bibinfo  {journal}
  {Rev. Mod. Phys.}\ }\textbf {\bibinfo {volume} {92}},\ \bibinfo {pages}
  {041001} (\bibinfo {year} {2020})}\BibitemShut {NoStop}%
\bibitem [{\citenamefont {Tobias}(2021)}]{Tobias21}%
  \BibitemOpen
  \bibfield  {author} {\bibinfo {author} {\bibfnamefont {S.~M.}\ \bibnamefont
  {Tobias}},\ }\href@noop {} {\bibfield  {journal} {\bibinfo  {journal} {J.
  Fluid Mech.}\ }\textbf {\bibinfo {volume} {912}},\ \bibinfo {pages} {P1}
  (\bibinfo {year} {2021})}\BibitemShut {NoStop}%
\bibitem [{\citenamefont {Landeau}\ \emph {et~al.}(2022)\citenamefont
  {Landeau}, \citenamefont {Fournier}, \citenamefont {Nataf}, \citenamefont
  {Cébron},\ and\ \citenamefont {Schaeffer}}]{Landeau22}%
  \BibitemOpen
  \bibfield  {author} {\bibinfo {author} {\bibfnamefont {M.}~\bibnamefont
  {Landeau}}, \bibinfo {author} {\bibfnamefont {A.}~\bibnamefont {Fournier}},
  \bibinfo {author} {\bibfnamefont {H.-C.}\ \bibnamefont {Nataf}}, \bibinfo
  {author} {\bibfnamefont {D.}~\bibnamefont {Cébron}}, \ and\ \bibinfo
  {author} {\bibfnamefont {N.}~\bibnamefont {Schaeffer}},\ }\href@noop {}
  {\bibfield  {journal} {\bibinfo  {journal} {Nat. Rev. Earth Environ.}\
  }\textbf {\bibinfo {volume} {3}},\ \bibinfo {pages} {255} (\bibinfo {year}
  {2022})}\BibitemShut {NoStop}%
\bibitem [{\citenamefont {Plumley}\ and\ \citenamefont
  {Julien}(2019)}]{Plumley19}%
  \BibitemOpen
  \bibfield  {author} {\bibinfo {author} {\bibfnamefont {M.}~\bibnamefont
  {Plumley}}\ and\ \bibinfo {author} {\bibfnamefont {K.}~\bibnamefont
  {Julien}},\ }\href@noop {} {\bibfield  {journal} {\bibinfo  {journal} {Earth
  Space Sci.}\ }\textbf {\bibinfo {volume} {34}},\ \bibinfo {pages}
  {1580–1592} (\bibinfo {year} {2019})}\BibitemShut {NoStop}%
\bibitem [{\citenamefont {Zhong}\ \emph {et~al.}(2009)\citenamefont {Zhong},
  \citenamefont {Stevens}, \citenamefont {Clercx}, \citenamefont {Verzicco},
  \citenamefont {Lohse},\ and\ \citenamefont {Ahlers}}]{Zhong09}%
  \BibitemOpen
  \bibfield  {author} {\bibinfo {author} {\bibfnamefont {J.-Q.}\ \bibnamefont
  {Zhong}}, \bibinfo {author} {\bibfnamefont {R.~J. A.~M.}\ \bibnamefont
  {Stevens}}, \bibinfo {author} {\bibfnamefont {H.~J.~H.}\ \bibnamefont
  {Clercx}}, \bibinfo {author} {\bibfnamefont {R.}~\bibnamefont {Verzicco}},
  \bibinfo {author} {\bibfnamefont {D.}~\bibnamefont {Lohse}}, \ and\ \bibinfo
  {author} {\bibfnamefont {G.}~\bibnamefont {Ahlers}},\ }\href@noop {}
  {\bibfield  {journal} {\bibinfo  {journal} {Phys. Rev. Lett.}\ }\textbf
  {\bibinfo {volume} {102}},\ \bibinfo {pages} {044502} (\bibinfo {year}
  {2009})}\BibitemShut {NoStop}%
\bibitem [{\citenamefont {King}\ \emph {et~al.}(2009)\citenamefont {King},
  \citenamefont {Stellmach}, \citenamefont {Noir}, \citenamefont {Hansen},\
  and\ \citenamefont {Aurnou}}]{King09}%
  \BibitemOpen
  \bibfield  {author} {\bibinfo {author} {\bibfnamefont {E.~M.}\ \bibnamefont
  {King}}, \bibinfo {author} {\bibfnamefont {S.}~\bibnamefont {Stellmach}},
  \bibinfo {author} {\bibfnamefont {J.}~\bibnamefont {Noir}}, \bibinfo {author}
  {\bibfnamefont {U.}~\bibnamefont {Hansen}}, \ and\ \bibinfo {author}
  {\bibfnamefont {J.~M.}\ \bibnamefont {Aurnou}},\ }\href@noop {} {\bibfield
  {journal} {\bibinfo  {journal} {Nature}\ }\textbf {\bibinfo {volume} {457}},\
  \bibinfo {pages} {301–304} (\bibinfo {year} {2009})}\BibitemShut {NoStop}%
\bibitem [{\citenamefont {Stevens}\ \emph {et~al.}(2009)\citenamefont
  {Stevens}, \citenamefont {Zhong}, \citenamefont {Clercx}, \citenamefont
  {Ahlers},\ and\ \citenamefont {Lohse}}]{Stevens09}%
  \BibitemOpen
  \bibfield  {author} {\bibinfo {author} {\bibfnamefont {R.~J. A.~M.}\
  \bibnamefont {Stevens}}, \bibinfo {author} {\bibfnamefont {J.-Q.}\
  \bibnamefont {Zhong}}, \bibinfo {author} {\bibfnamefont {H.~J.~H.}\
  \bibnamefont {Clercx}}, \bibinfo {author} {\bibfnamefont {G.}~\bibnamefont
  {Ahlers}}, \ and\ \bibinfo {author} {\bibfnamefont {D.}~\bibnamefont
  {Lohse}},\ }\href@noop {} {\bibfield  {journal} {\bibinfo  {journal} {Phys.
  Rev. Lett.}\ }\textbf {\bibinfo {volume} {103}},\ \bibinfo {pages} {024503}
  (\bibinfo {year} {2009})}\BibitemShut {NoStop}%
\bibitem [{\citenamefont {Stevens}\ \emph {et~al.}(2010)\citenamefont
  {Stevens}, \citenamefont {Clercx},\ and\ \citenamefont {Lohse}}]{Stevens10}%
  \BibitemOpen
  \bibfield  {author} {\bibinfo {author} {\bibfnamefont {R.~J. A.~M.}\
  \bibnamefont {Stevens}}, \bibinfo {author} {\bibfnamefont {H.~J.~H.}\
  \bibnamefont {Clercx}}, \ and\ \bibinfo {author} {\bibfnamefont
  {D.}~\bibnamefont {Lohse}},\ }\href@noop {} {\bibfield  {journal} {\bibinfo
  {journal} {New J. Phys.}\ }\textbf {\bibinfo {volume} {12}},\ \bibinfo
  {pages} {075005} (\bibinfo {year} {2010})}\BibitemShut {NoStop}%
\bibitem [{\citenamefont {Stevens}\ \emph {et~al.}(2013)\citenamefont
  {Stevens}, \citenamefont {Clercx},\ and\ \citenamefont {Lohse}}]{Stevens13}%
  \BibitemOpen
  \bibfield  {author} {\bibinfo {author} {\bibfnamefont {R.~J. A.~M.}\
  \bibnamefont {Stevens}}, \bibinfo {author} {\bibfnamefont {H.~J.~H.}\
  \bibnamefont {Clercx}}, \ and\ \bibinfo {author} {\bibfnamefont
  {D.}~\bibnamefont {Lohse}},\ }\href@noop {} {\bibfield  {journal} {\bibinfo
  {journal} {Eur. J. of Mech.}\ }\textbf {\bibinfo {volume} {40}},\ \bibinfo
  {pages} {41} (\bibinfo {year} {2013})}\BibitemShut {NoStop}%
\bibitem [{\citenamefont {Ecke}\ and\ \citenamefont {Niemela}(2014)}]{Ecke14}%
  \BibitemOpen
  \bibfield  {author} {\bibinfo {author} {\bibfnamefont {R.~E.}\ \bibnamefont
  {Ecke}}\ and\ \bibinfo {author} {\bibfnamefont {J.~J.}\ \bibnamefont
  {Niemela}},\ }\href@noop {} {\bibfield  {journal} {\bibinfo  {journal} {Phys.
  Rev. Lett.}\ }\textbf {\bibinfo {volume} {113}},\ \bibinfo {pages} {114301}
  (\bibinfo {year} {2014})}\BibitemShut {NoStop}%
\bibitem [{\citenamefont {Stellmach}\ \emph {et~al.}(2014)\citenamefont
  {Stellmach}, \citenamefont {Lischper}, \citenamefont {Julien}, \citenamefont
  {Vasil}, \citenamefont {Cheng}, \citenamefont {Ribeiro}, \citenamefont
  {King},\ and\ \citenamefont {Aurnou}}]{Stellmach14}%
  \BibitemOpen
  \bibfield  {author} {\bibinfo {author} {\bibfnamefont {S.}~\bibnamefont
  {Stellmach}}, \bibinfo {author} {\bibfnamefont {M.}~\bibnamefont {Lischper}},
  \bibinfo {author} {\bibfnamefont {K.}~\bibnamefont {Julien}}, \bibinfo
  {author} {\bibfnamefont {G.}~\bibnamefont {Vasil}}, \bibinfo {author}
  {\bibfnamefont {J.~S.}\ \bibnamefont {Cheng}}, \bibinfo {author}
  {\bibfnamefont {A.}~\bibnamefont {Ribeiro}}, \bibinfo {author} {\bibfnamefont
  {E.~M.}\ \bibnamefont {King}}, \ and\ \bibinfo {author} {\bibfnamefont
  {J.~M.}\ \bibnamefont {Aurnou}},\ }\href@noop {} {\bibfield  {journal}
  {\bibinfo  {journal} {Phys. Rev. Lett.}\ }\textbf {\bibinfo {volume} {113}},\
  \bibinfo {pages} {254501} (\bibinfo {year} {2014})}\BibitemShut {NoStop}%
\bibitem [{\citenamefont {Horn}\ and\ \citenamefont
  {Shishkina}(2015)}]{Horn2015}%
  \BibitemOpen
  \bibfield  {author} {\bibinfo {author} {\bibfnamefont {S.}~\bibnamefont
  {Horn}}\ and\ \bibinfo {author} {\bibfnamefont {O.}~\bibnamefont
  {Shishkina}},\ }\href@noop {} {\bibfield  {journal} {\bibinfo  {journal} {J.
  Fluid Mech.}\ }\textbf {\bibinfo {volume} {762}},\ \bibinfo {pages} {232}
  (\bibinfo {year} {2015})}\BibitemShut {NoStop}%
\bibitem [{\citenamefont {Weiss}\ \emph {et~al.}(2016)\citenamefont {Weiss},
  \citenamefont {Wei},\ and\ \citenamefont {Ahlers}}]{Weiss16}%
  \BibitemOpen
  \bibfield  {author} {\bibinfo {author} {\bibfnamefont {S.}~\bibnamefont
  {Weiss}}, \bibinfo {author} {\bibfnamefont {P.}~\bibnamefont {Wei}}, \ and\
  \bibinfo {author} {\bibfnamefont {G.}~\bibnamefont {Ahlers}},\ }\href@noop {}
  {\bibfield  {journal} {\bibinfo  {journal} {Phys. Rev. E}\ }\textbf {\bibinfo
  {volume} {03}},\ \bibinfo {pages} {043102} (\bibinfo {year}
  {2016})}\BibitemShut {NoStop}%
\bibitem [{\citenamefont {Plumley}\ \emph {et~al.}(2017)\citenamefont
  {Plumley}, \citenamefont {Julien}, \citenamefont {Marti},\ and\ \citenamefont
  {Stellmach}}]{Plumley17}%
  \BibitemOpen
  \bibfield  {author} {\bibinfo {author} {\bibfnamefont {M.}~\bibnamefont
  {Plumley}}, \bibinfo {author} {\bibfnamefont {K.}~\bibnamefont {Julien}},
  \bibinfo {author} {\bibfnamefont {P.}~\bibnamefont {Marti}}, \ and\ \bibinfo
  {author} {\bibfnamefont {S.}~\bibnamefont {Stellmach}},\ }\href@noop {}
  {\bibfield  {journal} {\bibinfo  {journal} {Phys. Rev. Fluids}\ }\textbf
  {\bibinfo {volume} {2}},\ \bibinfo {pages} {094801} (\bibinfo {year}
  {2017})}\BibitemShut {NoStop}%
\bibitem [{\citenamefont {Zhang}\ \emph {et~al.}(2020)\citenamefont {Zhang},
  \citenamefont {van Gils}, \citenamefont {Horn}, \citenamefont {Wedi},
  \citenamefont {Zwirner}, \citenamefont {Ahlers}, \citenamefont {Ecke},
  \citenamefont {Weiss}, \citenamefont {Bodenschatz},\ and\ \citenamefont
  {Shishkina}}]{Zhang20}%
  \BibitemOpen
  \bibfield  {author} {\bibinfo {author} {\bibfnamefont {X.}~\bibnamefont
  {Zhang}}, \bibinfo {author} {\bibfnamefont {D.~P.~M.}\ \bibnamefont {van
  Gils}}, \bibinfo {author} {\bibfnamefont {S.}~\bibnamefont {Horn}}, \bibinfo
  {author} {\bibfnamefont {M.}~\bibnamefont {Wedi}}, \bibinfo {author}
  {\bibfnamefont {L.}~\bibnamefont {Zwirner}}, \bibinfo {author} {\bibfnamefont
  {G.}~\bibnamefont {Ahlers}}, \bibinfo {author} {\bibfnamefont {R.~E.}\
  \bibnamefont {Ecke}}, \bibinfo {author} {\bibfnamefont {S.}~\bibnamefont
  {Weiss}}, \bibinfo {author} {\bibfnamefont {E.}~\bibnamefont {Bodenschatz}},
  \ and\ \bibinfo {author} {\bibfnamefont {O.}~\bibnamefont {Shishkina}},\
  }\href@noop {} {\bibfield  {journal} {\bibinfo  {journal} {Phys. Rev. Lett.}\
  }\textbf {\bibinfo {volume} {124}},\ \bibinfo {pages} {084505} (\bibinfo
  {year} {2020})}\BibitemShut {NoStop}%
\bibitem [{\citenamefont {Aurnou}\ \emph {et~al.}(2020)\citenamefont {Aurnou},
  \citenamefont {Horn},\ and\ \citenamefont {Julien}}]{Aurnou20}%
  \BibitemOpen
  \bibfield  {author} {\bibinfo {author} {\bibfnamefont {J.~M.}\ \bibnamefont
  {Aurnou}}, \bibinfo {author} {\bibfnamefont {S.}~\bibnamefont {Horn}}, \ and\
  \bibinfo {author} {\bibfnamefont {K.}~\bibnamefont {Julien}},\ }\href@noop {}
  {\bibfield  {journal} {\bibinfo  {journal} {Phys. Rev. Res.}\ }\textbf
  {\bibinfo {volume} {2}},\ \bibinfo {pages} {043115} (\bibinfo {year}
  {2020})}\BibitemShut {NoStop}%
\bibitem [{\citenamefont {Cheng}\ \emph {et~al.}(2020)\citenamefont {Cheng},
  \citenamefont {Madonia}, \citenamefont {{Aguirre Guzm\'an}},\ and\
  \citenamefont {Kunnen}}]{Cheng20}%
  \BibitemOpen
  \bibfield  {author} {\bibinfo {author} {\bibfnamefont {J.~S.}\ \bibnamefont
  {Cheng}}, \bibinfo {author} {\bibfnamefont {M.}~\bibnamefont {Madonia}},
  \bibinfo {author} {\bibfnamefont {A.~J.}\ \bibnamefont {{Aguirre Guzm\'an}}},
  \ and\ \bibinfo {author} {\bibfnamefont {R.~P.~J.}\ \bibnamefont {Kunnen}},\
  }\href@noop {} {\bibfield  {journal} {\bibinfo  {journal} {Phys. Rev.
  Fluids}\ }\textbf {\bibinfo {volume} {5}},\ \bibinfo {pages} {113501}
  (\bibinfo {year} {2020})}\BibitemShut {NoStop}%
\bibitem [{\citenamefont {{Aguirre Guzm\'an}}\ \emph
  {et~al.}(2020)\citenamefont {{Aguirre Guzm\'an}}, \citenamefont {Madonia},
  \citenamefont {Cheng}, \citenamefont {{Ostilla-M{ó}nico}}, \citenamefont
  {Clercx},\ and\ \citenamefont {Kunnen}}]{Guzman20}%
  \BibitemOpen
  \bibfield  {author} {\bibinfo {author} {\bibfnamefont {A.~J.}\ \bibnamefont
  {{Aguirre Guzm\'an}}}, \bibinfo {author} {\bibfnamefont {M.}~\bibnamefont
  {Madonia}}, \bibinfo {author} {\bibfnamefont {J.~S.}\ \bibnamefont {Cheng}},
  \bibinfo {author} {\bibfnamefont {R.}~\bibnamefont {{Ostilla-M{ó}nico}}},
  \bibinfo {author} {\bibfnamefont {H.~J.~H.}\ \bibnamefont {Clercx}}, \ and\
  \bibinfo {author} {\bibfnamefont {R.~P.~J.}\ \bibnamefont {Kunnen}},\
  }\href@noop {} {\bibfield  {journal} {\bibinfo  {journal} {Phys. Rev. Lett.}\
  }\textbf {\bibinfo {volume} {125}},\ \bibinfo {pages} {214501} (\bibinfo
  {year} {2020})}\BibitemShut {NoStop}%
\bibitem [{\citenamefont {Jiang}\ \emph {et~al.}(2020)\citenamefont {Jiang},
  \citenamefont {Zhu}, \citenamefont {Wang}, \citenamefont {Huisman},\ and\
  \citenamefont {Sun}}]{Jiang20}%
  \BibitemOpen
  \bibfield  {author} {\bibinfo {author} {\bibfnamefont {H.}~\bibnamefont
  {Jiang}}, \bibinfo {author} {\bibfnamefont {X.}~\bibnamefont {Zhu}}, \bibinfo
  {author} {\bibfnamefont {D.}~\bibnamefont {Wang}}, \bibinfo {author}
  {\bibfnamefont {S.~G.}\ \bibnamefont {Huisman}}, \ and\ \bibinfo {author}
  {\bibfnamefont {C.}~\bibnamefont {Sun}},\ }\href@noop {} {\bibfield
  {journal} {\bibinfo  {journal} {Sci. Adv.}\ }\textbf {\bibinfo {volume}
  {6}},\ \bibinfo {pages} {eabb8676} (\bibinfo {year} {2020})}\BibitemShut
  {NoStop}%
\bibitem [{\citenamefont {Bouillaut}\ \emph {et~al.}(2021)\citenamefont
  {Bouillaut}, \citenamefont {Miquel}, \citenamefont {Julien}, \citenamefont
  {{Aumaître}},\ and\ \citenamefont {Gallet}}]{Bouillaut21}%
  \BibitemOpen
  \bibfield  {author} {\bibinfo {author} {\bibfnamefont {V.}~\bibnamefont
  {Bouillaut}}, \bibinfo {author} {\bibfnamefont {B.}~\bibnamefont {Miquel}},
  \bibinfo {author} {\bibfnamefont {K.}~\bibnamefont {Julien}}, \bibinfo
  {author} {\bibfnamefont {S.}~\bibnamefont {{Aumaître}}}, \ and\ \bibinfo
  {author} {\bibfnamefont {B.}~\bibnamefont {Gallet}},\ }\href@noop {}
  {\bibfield  {journal} {\bibinfo  {journal} {Proc. Natl. Acad. Sci. U.S.A.}\
  }\textbf {\bibinfo {volume} {118}},\ \bibinfo {pages} {44} (\bibinfo {year}
  {2021})}\BibitemShut {NoStop}%
\bibitem [{\citenamefont {Lu}\ \emph {et~al.}(2021)\citenamefont {Lu},
  \citenamefont {Ding}, \citenamefont {Shi}, \citenamefont {Xia},\ and\
  \citenamefont {Zhong}}]{Lu21}%
  \BibitemOpen
  \bibfield  {author} {\bibinfo {author} {\bibfnamefont {H.}~\bibnamefont
  {Lu}}, \bibinfo {author} {\bibfnamefont {G.}~\bibnamefont {Ding}}, \bibinfo
  {author} {\bibfnamefont {J.}~\bibnamefont {Shi}}, \bibinfo {author}
  {\bibfnamefont {K.}~\bibnamefont {Xia}}, \ and\ \bibinfo {author}
  {\bibfnamefont {J.}~\bibnamefont {Zhong}},\ }\href@noop {} {\bibfield
  {journal} {\bibinfo  {journal} {Phys. Rev. Fluids}\ }\textbf {\bibinfo
  {volume} {6}},\ \bibinfo {pages} {L071501} (\bibinfo {year}
  {2021})}\BibitemShut {NoStop}%
\bibitem [{Sup()}]{Supp}%
  \BibitemOpen
  \href@noop {} {\bibinfo  {journal} {See Supplemental Material at ...}\ ,\
  \bibinfo {pages} {for two videos on the flow structures in the two regimes,
  and the details on the conducted DNS.}}\BibitemShut {Stop}%
\bibitem [{\citenamefont {Stevenson}(1979)}]{Stevenson79}%
  \BibitemOpen
\bibfield  {journal} {  }\bibfield  {author} {\bibinfo {author} {\bibfnamefont
  {D.~J.}\ \bibnamefont {Stevenson}},\ }\href@noop {} {\bibfield  {journal}
  {\bibinfo  {journal} {Geophys. Astrophys. Fluid Dyn.}\ }\textbf {\bibinfo
  {volume} {12}},\ \bibinfo {pages} {139} (\bibinfo {year} {1979})}\BibitemShut
  {NoStop}%
\bibitem [{\citenamefont {Gillet}\ and\ \citenamefont
  {Jones}(2006)}]{Gillet06}%
  \BibitemOpen
  \bibfield  {author} {\bibinfo {author} {\bibfnamefont {N.}~\bibnamefont
  {Gillet}}\ and\ \bibinfo {author} {\bibfnamefont {C.~A.}\ \bibnamefont
  {Jones}},\ }\href@noop {} {\bibfield  {journal} {\bibinfo  {journal} {J.
  Fluid Mech.}\ }\textbf {\bibinfo {volume} {554}},\ \bibinfo {pages} {343}
  (\bibinfo {year} {2006})}\BibitemShut {NoStop}%
\bibitem [{\citenamefont {Julien}\ \emph
  {et~al.}(2012{\natexlab{a}})\citenamefont {Julien}, \citenamefont {Knobloch},
  \citenamefont {Rubio},\ and\ \citenamefont {Vasil}}]{Julien12}%
  \BibitemOpen
  \bibfield  {author} {\bibinfo {author} {\bibfnamefont {K.}~\bibnamefont
  {Julien}}, \bibinfo {author} {\bibfnamefont {E.}~\bibnamefont {Knobloch}},
  \bibinfo {author} {\bibfnamefont {A.~M.}\ \bibnamefont {Rubio}}, \ and\
  \bibinfo {author} {\bibfnamefont {G.~M.}\ \bibnamefont {Vasil}},\ }\href@noop
  {} {\bibfield  {journal} {\bibinfo  {journal} {Phys. Rev. Lett.}\ }\textbf
  {\bibinfo {volume} {109}},\ \bibinfo {pages} {254503} (\bibinfo {year}
  {2012}{\natexlab{a}})}\BibitemShut {NoStop}%
\bibitem [{\citenamefont {Julien}\ \emph
  {et~al.}(2012{\natexlab{b}})\citenamefont {Julien}, \citenamefont {Rubio},
  \citenamefont {Grooms},\ and\ \citenamefont {Knobloch}}]{Julien12a}%
  \BibitemOpen
  \bibfield  {author} {\bibinfo {author} {\bibfnamefont {K.}~\bibnamefont
  {Julien}}, \bibinfo {author} {\bibfnamefont {A.~M.}\ \bibnamefont {Rubio}},
  \bibinfo {author} {\bibfnamefont {I.}~\bibnamefont {Grooms}}, \ and\ \bibinfo
  {author} {\bibfnamefont {E.}~\bibnamefont {Knobloch}},\ }\href@noop {}
  {\bibfield  {journal} {\bibinfo  {journal} {Geophys. Astrophys. Fluid Dyn.}\
  }\textbf {\bibinfo {volume} {106}},\ \bibinfo {pages} {392} (\bibinfo {year}
  {2012}{\natexlab{b}})}\BibitemShut {NoStop}%
\bibitem [{\citenamefont {Gastine}\ \emph {et~al.}(2016)\citenamefont
  {Gastine}, \citenamefont {Wicht},\ and\ \citenamefont {Aubert}}]{Gastine16}%
  \BibitemOpen
  \bibfield  {author} {\bibinfo {author} {\bibfnamefont {T.}~\bibnamefont
  {Gastine}}, \bibinfo {author} {\bibfnamefont {J.}~\bibnamefont {Wicht}}, \
  and\ \bibinfo {author} {\bibfnamefont {J.}~\bibnamefont {Aubert}},\
  }\href@noop {} {\bibfield  {journal} {\bibinfo  {journal} {J. Fluid Mech.}\
  }\textbf {\bibinfo {volume} {808}},\ \bibinfo {pages} {690–732} (\bibinfo
  {year} {2016})}\BibitemShut {NoStop}%
\bibitem [{\citenamefont {Schmitz}\ and\ \citenamefont
  {Tilgner}(2009)}]{Schmitz09}%
  \BibitemOpen
  \bibfield  {author} {\bibinfo {author} {\bibfnamefont {S.}~\bibnamefont
  {Schmitz}}\ and\ \bibinfo {author} {\bibfnamefont {A.}~\bibnamefont
  {Tilgner}},\ }\href@noop {} {\bibfield  {journal} {\bibinfo  {journal} {Phys.
  Rev. E}\ }\textbf {\bibinfo {volume} {80}},\ \bibinfo {pages} {0015305(R)}
  (\bibinfo {year} {2009})}\BibitemShut {NoStop}%
\bibitem [{\citenamefont {Wang}\ \emph {et~al.}(2021)\citenamefont {Wang},
  \citenamefont {Santelli}, \citenamefont {Lohse}, \citenamefont {Verzicco},\
  and\ \citenamefont {Stevens}}]{Wang21}%
  \BibitemOpen
  \bibfield  {author} {\bibinfo {author} {\bibfnamefont {G.}~\bibnamefont
  {Wang}}, \bibinfo {author} {\bibfnamefont {L.}~\bibnamefont {Santelli}},
  \bibinfo {author} {\bibfnamefont {D.}~\bibnamefont {Lohse}}, \bibinfo
  {author} {\bibfnamefont {R.}~\bibnamefont {Verzicco}}, \ and\ \bibinfo
  {author} {\bibfnamefont {R.~J. A.~M.}\ \bibnamefont {Stevens}},\ }\href@noop
  {} {\bibfield  {journal} {\bibinfo  {journal} {Geophys. Res. Lett.}\ }\textbf
  {\bibinfo {volume} {48}},\ \bibinfo {pages} {e2021GL095017.} (\bibinfo {year}
  {2021})}\BibitemShut {NoStop}%
\bibitem [{\citenamefont {Grooms}\ and\ \citenamefont
  {Whitehead}(2015)}]{Grooms2015}%
  \BibitemOpen
  \bibfield  {author} {\bibinfo {author} {\bibfnamefont {I.}~\bibnamefont
  {Grooms}}\ and\ \bibinfo {author} {\bibfnamefont {J.}~\bibnamefont
  {Whitehead}},\ }\href@noop {} {\bibfield  {journal} {\bibinfo  {journal}
  {Nonlinearity}\ }\textbf {\bibinfo {volume} {28}},\ \bibinfo {pages} {29}
  (\bibinfo {year} {2015})}\BibitemShut {NoStop}%
\bibitem [{\citenamefont {Malkus}(1954)}]{Malkus1954}%
  \BibitemOpen
  \bibfield  {author} {\bibinfo {author} {\bibfnamefont {M.~V.~R.}\
  \bibnamefont {Malkus}},\ }\href@noop {} {\bibfield  {journal} {\bibinfo
  {journal} {Proc. R. Soc. London A}\ }\textbf {\bibinfo {volume} {225}},\
  \bibinfo {pages} {196} (\bibinfo {year} {1954})}\BibitemShut {NoStop}%
\bibitem [{\citenamefont {Priestley}(1959)}]{Priestley1959}%
  \BibitemOpen
  \bibfield  {author} {\bibinfo {author} {\bibfnamefont {C.~H.~B.}\
  \bibnamefont {Priestley}},\ }\href@noop {} {\emph {\bibinfo {title}
  {{Turbulent Transfer in the Lower Atmosphere}}}}\ (\bibinfo  {publisher}
  {University of Chicago Press},\ \bibinfo {year} {1959})\BibitemShut {NoStop}%
\bibitem [{\citenamefont {Boubnov}\ and\ \citenamefont
  {Golitsyn}(1990)}]{Boubnov90}%
  \BibitemOpen
  \bibfield  {author} {\bibinfo {author} {\bibfnamefont {B.~M.}\ \bibnamefont
  {Boubnov}}\ and\ \bibinfo {author} {\bibfnamefont {G.~S.}\ \bibnamefont
  {Golitsyn}},\ }\href@noop {} {\bibfield  {journal} {\bibinfo  {journal} {J.
  of Fluid Mech.}\ }\textbf {\bibinfo {volume} {219}},\ \bibinfo {pages} {215}
  (\bibinfo {year} {1990})}\BibitemShut {NoStop}%
\bibitem [{\citenamefont {King}\ \emph {et~al.}(2012)\citenamefont {King},
  \citenamefont {Stellmach},\ and\ \citenamefont {Aurnou}}]{King12}%
  \BibitemOpen
  \bibfield  {author} {\bibinfo {author} {\bibfnamefont {E.~M.}\ \bibnamefont
  {King}}, \bibinfo {author} {\bibfnamefont {S.}~\bibnamefont {Stellmach}}, \
  and\ \bibinfo {author} {\bibfnamefont {J.~M.}\ \bibnamefont {Aurnou}},\
  }\href@noop {} {\bibfield  {journal} {\bibinfo  {journal} {J. Fluid Mech.}\
  }\textbf {\bibinfo {volume} {691}},\ \bibinfo {pages} {568–582} (\bibinfo
  {year} {2012})}\BibitemShut {NoStop}%
\bibitem [{\citenamefont {Cheng}\ and\ \citenamefont {Aurnou}(2016)}]{Cheng16}%
  \BibitemOpen
  \bibfield  {author} {\bibinfo {author} {\bibfnamefont {J.~S.}\ \bibnamefont
  {Cheng}}\ and\ \bibinfo {author} {\bibfnamefont {J.~M.}\ \bibnamefont
  {Aurnou}},\ }\href@noop {} {\bibfield  {journal} {\bibinfo  {journal} {Earth
  Planet Sci. Lett.}\ }\textbf {\bibinfo {volume} {436}},\ \bibinfo {pages}
  {121–129} (\bibinfo {year} {2016})}\BibitemShut {NoStop}%
\bibitem [{\citenamefont {Julien}\ \emph {et~al.}(2016)\citenamefont {Julien},
  \citenamefont {Aurnou}, \citenamefont {Calkins}, \citenamefont {Knobloch},
  \citenamefont {Marti}, \citenamefont {Stellmach},\ and\ \citenamefont
  {Vasil}}]{Julien16}%
  \BibitemOpen
  \bibfield  {author} {\bibinfo {author} {\bibfnamefont {K.}~\bibnamefont
  {Julien}}, \bibinfo {author} {\bibfnamefont {J.~M.}\ \bibnamefont {Aurnou}},
  \bibinfo {author} {\bibfnamefont {M.~A.}\ \bibnamefont {Calkins}}, \bibinfo
  {author} {\bibfnamefont {E.}~\bibnamefont {Knobloch}}, \bibinfo {author}
  {\bibfnamefont {P.}~\bibnamefont {Marti}}, \bibinfo {author} {\bibfnamefont
  {S.}~\bibnamefont {Stellmach}}, \ and\ \bibinfo {author} {\bibfnamefont
  {G.~M.}\ \bibnamefont {Vasil}},\ }\href@noop {} {\bibfield  {journal}
  {\bibinfo  {journal} {J. Fluid Mech.}\ }\textbf {\bibinfo {volume} {798}},\
  \bibinfo {pages} {50–87} (\bibinfo {year} {2016})}\BibitemShut {NoStop}%
\bibitem [{\citenamefont {{Aguirre Guzm\'an}}\ \emph
  {et~al.}(2021)\citenamefont {{Aguirre Guzm\'an}}, \citenamefont {Madonia},
  \citenamefont {Cheng}, \citenamefont {Ostilla-M\'onico}, \citenamefont
  {Clercx},\ and\ \citenamefont {Kunnen}}]{Guzman21}%
  \BibitemOpen
  \bibfield  {author} {\bibinfo {author} {\bibfnamefont {A.~J.}\ \bibnamefont
  {{Aguirre Guzm\'an}}}, \bibinfo {author} {\bibfnamefont {M.}~\bibnamefont
  {Madonia}}, \bibinfo {author} {\bibfnamefont {J.~S.}\ \bibnamefont {Cheng}},
  \bibinfo {author} {\bibfnamefont {R.}~\bibnamefont {Ostilla-M\'onico}},
  \bibinfo {author} {\bibfnamefont {H.~J.~H.}\ \bibnamefont {Clercx}}, \ and\
  \bibinfo {author} {\bibfnamefont {R.~P.~J.}\ \bibnamefont {Kunnen}},\
  }\href@noop {} {\bibfield  {journal} {\bibinfo  {journal} {J. Fluid Mech.}\
  }\textbf {\bibinfo {volume} {928}},\ \bibinfo {pages} {A16} (\bibinfo {year}
  {2021})}\BibitemShut {NoStop}%
\bibitem [{\citenamefont {Vorobieff}\ and\ \citenamefont
  {Ecke}(2002)}]{Vorobieff2002}%
  \BibitemOpen
  \bibfield  {author} {\bibinfo {author} {\bibfnamefont {P.}~\bibnamefont
  {Vorobieff}}\ and\ \bibinfo {author} {\bibfnamefont {R.~E.}\ \bibnamefont
  {Ecke}},\ }\href@noop {} {\bibfield  {journal} {\bibinfo  {journal} {J. Fluid
  Mech.}\ }\textbf {\bibinfo {volume} {458}},\ \bibinfo {pages} {191} (\bibinfo
  {year} {2002})}\BibitemShut {NoStop}%
\bibitem [{\citenamefont {Kunnen}\ \emph {et~al.}(2008)\citenamefont {Kunnen},
  \citenamefont {Clercx},\ and\ \citenamefont {Geurts}}]{Kunnen2008}%
  \BibitemOpen
  \bibfield  {author} {\bibinfo {author} {\bibfnamefont {R.~P.~J.}\
  \bibnamefont {Kunnen}}, \bibinfo {author} {\bibfnamefont {H.~J.~H.}\
  \bibnamefont {Clercx}}, \ and\ \bibinfo {author} {\bibfnamefont {B.~J.}\
  \bibnamefont {Geurts}},\ }\href@noop {} {\bibfield  {journal} {\bibinfo
  {journal} {Europhys. Lett.}\ }\textbf {\bibinfo {volume} {84}},\ \bibinfo
  {pages} {24001} (\bibinfo {year} {2008})}\BibitemShut {NoStop}%
\bibitem [{\citenamefont {Weiss}\ and\ \citenamefont
  {Ahlers}(2011{\natexlab{a}})}]{Weiss2011}%
  \BibitemOpen
  \bibfield  {author} {\bibinfo {author} {\bibfnamefont {S.}~\bibnamefont
  {Weiss}}\ and\ \bibinfo {author} {\bibfnamefont {G.}~\bibnamefont {Ahlers}},\
  }\href@noop {} {\bibfield  {journal} {\bibinfo  {journal} {J. Fluid Mech.}\
  }\textbf {\bibinfo {volume} {688}},\ \bibinfo {pages} {461} (\bibinfo {year}
  {2011}{\natexlab{a}})}\BibitemShut {NoStop}%
\bibitem [{\citenamefont {Weiss}\ and\ \citenamefont
  {Ahlers}(2011{\natexlab{b}})}]{Weiss2011b}%
  \BibitemOpen
  \bibfield  {author} {\bibinfo {author} {\bibfnamefont {S.}~\bibnamefont
  {Weiss}}\ and\ \bibinfo {author} {\bibfnamefont {G.}~\bibnamefont {Ahlers}},\
  }\href@noop {} {\bibfield  {journal} {\bibinfo  {journal} {J. Fluid Mech.}\
  }\textbf {\bibinfo {volume} {684}},\ \bibinfo {pages} {407} (\bibinfo {year}
  {2011}{\natexlab{b}})}\BibitemShut {NoStop}%
\bibitem [{\citenamefont {King}\ \emph {et~al.}(2013)\citenamefont {King},
  \citenamefont {Stellmach},\ and\ \citenamefont {Buffett}}]{King13}%
  \BibitemOpen
  \bibfield  {author} {\bibinfo {author} {\bibfnamefont {E.~M.}\ \bibnamefont
  {King}}, \bibinfo {author} {\bibfnamefont {S.}~\bibnamefont {Stellmach}}, \
  and\ \bibinfo {author} {\bibfnamefont {B.}~\bibnamefont {Buffett}},\
  }\href@noop {} {\bibfield  {journal} {\bibinfo  {journal} {J. Fluid Mech.}\
  }\textbf {\bibinfo {volume} {717}},\ \bibinfo {pages} {449} (\bibinfo {year}
  {2013})}\BibitemShut {NoStop}%
\bibitem [{\citenamefont {Horn}\ and\ \citenamefont
  {Shishkina}(2014)}]{Horn14}%
  \BibitemOpen
  \bibfield  {author} {\bibinfo {author} {\bibfnamefont {S.}~\bibnamefont
  {Horn}}\ and\ \bibinfo {author} {\bibfnamefont {O.}~\bibnamefont
  {Shishkina}},\ }\href@noop {} {\bibfield  {journal} {\bibinfo  {journal}
  {Phys. Fluids}\ }\textbf {\bibinfo {volume} {26}},\ \bibinfo {pages} {055111}
  (\bibinfo {year} {2014})}\BibitemShut {NoStop}%
\bibitem [{\citenamefont {Hartmann}\ \emph {et~al.}(2022)\citenamefont
  {Hartmann}, \citenamefont {Verzicco}, \citenamefont {{Klein Kranenbarg}},
  \citenamefont {Lohse},\ and\ \citenamefont {Stevens}}]{Hartmann22}%
  \BibitemOpen
  \bibfield  {author} {\bibinfo {author} {\bibfnamefont {R.}~\bibnamefont
  {Hartmann}}, \bibinfo {author} {\bibfnamefont {R.}~\bibnamefont {Verzicco}},
  \bibinfo {author} {\bibfnamefont {L.}~\bibnamefont {{Klein Kranenbarg}}},
  \bibinfo {author} {\bibfnamefont {D.}~\bibnamefont {Lohse}}, \ and\ \bibinfo
  {author} {\bibfnamefont {R.~J. A.~M.}\ \bibnamefont {Stevens}},\ }\href@noop
  {} {\bibfield  {journal} {\bibinfo  {journal} {J. Fluid Mech.}\ }\textbf
  {\bibinfo {volume} {939}},\ \bibinfo {pages} {A1} (\bibinfo {year}
  {2022})}\BibitemShut {NoStop}%
\bibitem [{\citenamefont {Landau}\ and\ \citenamefont
  {Lifshitz}(1987)}]{Landau1987}%
  \BibitemOpen
  \bibfield  {author} {\bibinfo {author} {\bibfnamefont {L.~D.}\ \bibnamefont
  {Landau}}\ and\ \bibinfo {author} {\bibfnamefont {E.~M.}\ \bibnamefont
  {Lifshitz}},\ }\href@noop {} {\emph {\bibinfo {title} {{Fluid Mechanics}}}},\
  \bibinfo {edition} {2nd}\ ed.,\ \bibinfo {series} {Course of Theoretical
  Physics}, Vol.~\bibinfo {volume} {6}\ (\bibinfo  {publisher} {Butterworth
  Heinemann},\ \bibinfo {year} {1987})\BibitemShut {NoStop}%
\bibitem [{\citenamefont {Sprague}\ \emph {et~al.}(2006)\citenamefont
  {Sprague}, \citenamefont {Julien}, \citenamefont {Knobloch},\ and\
  \citenamefont {Werne}}]{Sprague2006}%
  \BibitemOpen
  \bibfield  {author} {\bibinfo {author} {\bibfnamefont {M.}~\bibnamefont
  {Sprague}}, \bibinfo {author} {\bibfnamefont {K.}~\bibnamefont {Julien}},
  \bibinfo {author} {\bibfnamefont {E.}~\bibnamefont {Knobloch}}, \ and\
  \bibinfo {author} {\bibfnamefont {J.}~\bibnamefont {Werne}},\ }\href@noop {}
  {\bibfield  {journal} {\bibinfo  {journal} {J. Fluid Mech.}\ }\textbf
  {\bibinfo {volume} {551}},\ \bibinfo {pages} {141} (\bibinfo {year}
  {2006})}\BibitemShut {NoStop}%
\bibitem [{\citenamefont {Plumley}\ \emph {et~al.}(2016)\citenamefont
  {Plumley}, \citenamefont {Julien}, \citenamefont {Marti},\ and\ \citenamefont
  {Stellmach}}]{Plumley16}%
  \BibitemOpen
  \bibfield  {author} {\bibinfo {author} {\bibfnamefont {M.}~\bibnamefont
  {Plumley}}, \bibinfo {author} {\bibfnamefont {K.}~\bibnamefont {Julien}},
  \bibinfo {author} {\bibfnamefont {P.}~\bibnamefont {Marti}}, \ and\ \bibinfo
  {author} {\bibfnamefont {S.}~\bibnamefont {Stellmach}},\ }\href@noop {}
  {\bibfield  {journal} {\bibinfo  {journal} {J. Fluid Mech.}\ }\textbf
  {\bibinfo {volume} {803}},\ \bibinfo {pages} {51} (\bibinfo {year}
  {2016})}\BibitemShut {NoStop}%
\bibitem [{\citenamefont {Rossby}(1969)}]{Rossby1969}%
  \BibitemOpen
  \bibfield  {author} {\bibinfo {author} {\bibfnamefont {T.~H.}\ \bibnamefont
  {Rossby}},\ }\href@noop {} {\bibfield  {journal} {\bibinfo  {journal} {J.
  Fluid Mech.}\ }\textbf {\bibinfo {volume} {36}},\ \bibinfo {pages} {309}
  (\bibinfo {year} {1969})}\BibitemShut {NoStop}%
\bibitem [{\citenamefont {Ecke}\ \emph {et~al.}(1992)\citenamefont {Ecke},
  \citenamefont {Zhong},\ and\ \citenamefont {Knobloch}}]{Ecke1992}%
  \BibitemOpen
  \bibfield  {author} {\bibinfo {author} {\bibfnamefont {R.}~\bibnamefont
  {Ecke}}, \bibinfo {author} {\bibfnamefont {F.}~\bibnamefont {Zhong}}, \ and\
  \bibinfo {author} {\bibfnamefont {E.}~\bibnamefont {Knobloch}},\ }\href@noop
  {} {\bibfield  {journal} {\bibinfo  {journal} {Europhys. Lett.}\ }\textbf
  {\bibinfo {volume} {19}},\ \bibinfo {pages} {177} (\bibinfo {year}
  {1992})}\BibitemShut {NoStop}%
\bibitem [{\citenamefont {Goldstein}\ \emph {et~al.}(1993)\citenamefont
  {Goldstein}, \citenamefont {Knobloch}, \citenamefont {Mercader},\ and\
  \citenamefont {Net}}]{Goldstein1993}%
  \BibitemOpen
  \bibfield  {author} {\bibinfo {author} {\bibfnamefont {H.~F.}\ \bibnamefont
  {Goldstein}}, \bibinfo {author} {\bibfnamefont {E.}~\bibnamefont {Knobloch}},
  \bibinfo {author} {\bibfnamefont {I.}~\bibnamefont {Mercader}}, \ and\
  \bibinfo {author} {\bibfnamefont {M.}~\bibnamefont {Net}},\ }\href@noop {}
  {\bibfield  {journal} {\bibinfo  {journal} {J. Fluid Mech.}\ }\textbf
  {\bibinfo {volume} {248}},\ \bibinfo {pages} {583} (\bibinfo {year}
  {1993})}\BibitemShut {NoStop}%
\bibitem [{\citenamefont {Favier}\ and\ \citenamefont
  {Knobloch}(2020)}]{Favier2020}%
  \BibitemOpen
  \bibfield  {author} {\bibinfo {author} {\bibfnamefont {B.}~\bibnamefont
  {Favier}}\ and\ \bibinfo {author} {\bibfnamefont {E.}~\bibnamefont
  {Knobloch}},\ }\href@noop {} {\bibfield  {journal} {\bibinfo  {journal} {J.
  Fluid Mech.}\ }\textbf {\bibinfo {volume} {895}},\ \bibinfo {pages} {R1}
  (\bibinfo {year} {2020})}\BibitemShut {NoStop}%
\bibitem [{\citenamefont {Shishkina}(2020)}]{Shishkina2020}%
  \BibitemOpen
  \bibfield  {author} {\bibinfo {author} {\bibfnamefont {O.}~\bibnamefont
  {Shishkina}},\ }\href@noop {} {\bibfield  {journal} {\bibinfo  {journal} {J.
  Fluid Mech.}\ }\textbf {\bibinfo {volume} {898}},\ \bibinfo {pages} {F1}
  (\bibinfo {year} {2020})}\BibitemShut {NoStop}%
\bibitem [{\citenamefont {de~Wit}\ \emph {et~al.}(2020)\citenamefont {de~Wit},
  \citenamefont {{Aguirre Guzm\'an}}, \citenamefont {Madonia}, \citenamefont
  {Cheng}, \citenamefont {Clercx},\ and\ \citenamefont {Kunnen}}]{Wit2020}%
  \BibitemOpen
  \bibfield  {author} {\bibinfo {author} {\bibfnamefont {X.~M.}\ \bibnamefont
  {de~Wit}}, \bibinfo {author} {\bibfnamefont {A.~J.}\ \bibnamefont {{Aguirre
  Guzm\'an}}}, \bibinfo {author} {\bibfnamefont {M.}~\bibnamefont {Madonia}},
  \bibinfo {author} {\bibfnamefont {J.~S.}\ \bibnamefont {Cheng}}, \bibinfo
  {author} {\bibfnamefont {H.~J.~H.}\ \bibnamefont {Clercx}}, \ and\ \bibinfo
  {author} {\bibfnamefont {R.~P.~J.}\ \bibnamefont {Kunnen}},\ }\href@noop {}
  {\bibfield  {journal} {\bibinfo  {journal} {Phys. Rev. Fluids}\ }\textbf
  {\bibinfo {volume} {5}},\ \bibinfo {pages} {023502} (\bibinfo {year}
  {2020})}\BibitemShut {NoStop}%
\bibitem [{\citenamefont {Zhang}\ \emph {et~al.}(2021)\citenamefont {Zhang},
  \citenamefont {Ecke},\ and\ \citenamefont {Shishkina}}]{Zhang21}%
  \BibitemOpen
  \bibfield  {author} {\bibinfo {author} {\bibfnamefont {X.}~\bibnamefont
  {Zhang}}, \bibinfo {author} {\bibfnamefont {R.~E.}\ \bibnamefont {Ecke}}, \
  and\ \bibinfo {author} {\bibfnamefont {O.}~\bibnamefont {Shishkina}},\
  }\href@noop {} {\bibfield  {journal} {\bibinfo  {journal} {J. Fluid Mech.}\
  }\textbf {\bibinfo {volume} {915}},\ \bibinfo {pages} {A62} (\bibinfo {year}
  {2021})}\BibitemShut {NoStop}%
\bibitem [{\citenamefont {Ecke}\ \emph {et~al.}(2022)\citenamefont {Ecke},
  \citenamefont {Zhang},\ and\ \citenamefont {Shishkina}}]{Ecke22}%
  \BibitemOpen
  \bibfield  {author} {\bibinfo {author} {\bibfnamefont {R.~E.}\ \bibnamefont
  {Ecke}}, \bibinfo {author} {\bibfnamefont {X.}~\bibnamefont {Zhang}}, \ and\
  \bibinfo {author} {\bibfnamefont {O.}~\bibnamefont {Shishkina}},\ }\href@noop
  {} {\bibfield  {journal} {\bibinfo  {journal} {Phys. Rev. Fluids}\ }\textbf
  {\bibinfo {volume} {7}},\ \bibinfo {pages} {L011501} (\bibinfo {year}
  {2022})}\BibitemShut {NoStop}%
\bibitem [{\citenamefont {Wedi}\ \emph {et~al.}(2022)\citenamefont {Wedi},
  \citenamefont {M.Moturi}, \citenamefont {Funfschilling},\ and\ \citenamefont
  {Weiss}}]{Wedi22}%
  \BibitemOpen
  \bibfield  {author} {\bibinfo {author} {\bibfnamefont {M.}~\bibnamefont
  {Wedi}}, \bibinfo {author} {\bibfnamefont {V.}~\bibnamefont {M.Moturi}},
  \bibinfo {author} {\bibfnamefont {D.}~\bibnamefont {Funfschilling}}, \ and\
  \bibinfo {author} {\bibfnamefont {S.}~\bibnamefont {Weiss}},\ }\href@noop {}
  {\bibfield  {journal} {\bibinfo  {journal} {J. Fluid Mech.}\ }\textbf
  {\bibinfo {volume} {939}},\ \bibinfo {pages} {A14} (\bibinfo {year}
  {2022})}\BibitemShut {NoStop}%
\bibitem [{\citenamefont {Guervilly}\ \emph {et~al.}(2014)\citenamefont
  {Guervilly}, \citenamefont {Hughes},\ and\ \citenamefont
  {Jones}}]{Guervilly14}%
  \BibitemOpen
  \bibfield  {author} {\bibinfo {author} {\bibfnamefont {C.}~\bibnamefont
  {Guervilly}}, \bibinfo {author} {\bibfnamefont {D.~W.}\ \bibnamefont
  {Hughes}}, \ and\ \bibinfo {author} {\bibfnamefont {C.~A.}\ \bibnamefont
  {Jones}},\ }\href@noop {} {\bibfield  {journal} {\bibinfo  {journal} {J.
  Fluid Mech.}\ }\textbf {\bibinfo {volume} {758}},\ \bibinfo {pages}
  {407–435} (\bibinfo {year} {2014})}\BibitemShut {NoStop}%
\bibitem [{\citenamefont {Maffei}\ \emph {et~al.}(2021)\citenamefont {Maffei},
  \citenamefont {Krouss}, \citenamefont {Julien},\ and\ \citenamefont
  {Calkins}}]{Maffei21}%
  \BibitemOpen
  \bibfield  {author} {\bibinfo {author} {\bibfnamefont {S.}~\bibnamefont
  {Maffei}}, \bibinfo {author} {\bibfnamefont {M.~J.}\ \bibnamefont {Krouss}},
  \bibinfo {author} {\bibfnamefont {K.}~\bibnamefont {Julien}}, \ and\ \bibinfo
  {author} {\bibfnamefont {M.~A.}\ \bibnamefont {Calkins}},\ }\href@noop {}
  {\bibfield  {journal} {\bibinfo  {journal} {J. Fluid Mech.}\ }\textbf
  {\bibinfo {volume} {913}},\ \bibinfo {pages} {A18} (\bibinfo {year}
  {2021})}\BibitemShut {NoStop}%
\bibitem [{\citenamefont {Guervilly}\ and\ \citenamefont
  {Hughes}(2017)}]{Guervilly17}%
  \BibitemOpen
  \bibfield  {author} {\bibinfo {author} {\bibfnamefont {C.}~\bibnamefont
  {Guervilly}}\ and\ \bibinfo {author} {\bibfnamefont {D.~W.}\ \bibnamefont
  {Hughes}},\ }\href@noop {} {\bibfield  {journal} {\bibinfo  {journal} {Phys.
  Rev. Fluids}\ }\textbf {\bibinfo {volume} {2}},\ \bibinfo {pages} {113503}
  (\bibinfo {year} {2017})}\BibitemShut {NoStop}%
\bibitem [{\citenamefont {Julien}\ \emph {et~al.}(2018)\citenamefont {Julien},
  \citenamefont {Knobloch},\ and\ \citenamefont {Plumley}}]{Julien18}%
  \BibitemOpen
  \bibfield  {author} {\bibinfo {author} {\bibfnamefont {K.}~\bibnamefont
  {Julien}}, \bibinfo {author} {\bibfnamefont {E.}~\bibnamefont {Knobloch}}, \
  and\ \bibinfo {author} {\bibfnamefont {M.}~\bibnamefont {Plumley}},\
  }\href@noop {} {\bibfield  {journal} {\bibinfo  {journal} {J. Fluid Mech.}\
  }\textbf {\bibinfo {volume} {837}},\ \bibinfo {pages} {R4} (\bibinfo {year}
  {2018})}\BibitemShut {NoStop}%
\bibitem [{\citenamefont {Favier}\ \emph {et~al.}(2019)\citenamefont {Favier},
  \citenamefont {Guervilly},\ and\ \citenamefont {Knobloch}}]{Favier19}%
  \BibitemOpen
  \bibfield  {author} {\bibinfo {author} {\bibfnamefont {B.}~\bibnamefont
  {Favier}}, \bibinfo {author} {\bibfnamefont {C.}~\bibnamefont {Guervilly}}, \
  and\ \bibinfo {author} {\bibfnamefont {E.}~\bibnamefont {Knobloch}},\
  }\href@noop {} {\bibfield  {journal} {\bibinfo  {journal} {J. Fluid Mech.}\
  }\textbf {\bibinfo {volume} {864}},\ \bibinfo {pages} {R1} (\bibinfo {year}
  {2019})}\BibitemShut {NoStop}%
\bibitem [{\citenamefont {de~Wit}\ \emph {et~al.}(2022)\citenamefont {de~Wit},
  \citenamefont {{Aguirre Guzm\'an}}, \citenamefont {Madonia}, \citenamefont
  {Cheng}, \citenamefont {Clercx},\ and\ \citenamefont {Kunnen}}]{Wit22}%
  \BibitemOpen
  \bibfield  {author} {\bibinfo {author} {\bibfnamefont {X.~M.}\ \bibnamefont
  {de~Wit}}, \bibinfo {author} {\bibfnamefont {A.~J.}\ \bibnamefont {{Aguirre
  Guzm\'an}}}, \bibinfo {author} {\bibfnamefont {M.}~\bibnamefont {Madonia}},
  \bibinfo {author} {\bibfnamefont {J.~S.}\ \bibnamefont {Cheng}}, \bibinfo
  {author} {\bibfnamefont {H.~J.~H.}\ \bibnamefont {Clercx}}, \ and\ \bibinfo
  {author} {\bibfnamefont {R.~P.~J.}\ \bibnamefont {Kunnen}},\ }\href@noop {}
  {\bibfield  {journal} {\bibinfo  {journal} {J. Fluid Mech.}\ }\textbf
  {\bibinfo {volume} {963}},\ \bibinfo {pages} {A43} (\bibinfo {year}
  {2022})}\BibitemShut {NoStop}%
\end{thebibliography}%

\end{document}